\documentstyle[aps,graphicx,floats,tighten,preprint]{revtex}

\def\be{\begin{equation}}
\def\ee{\end{equation}}
\def\bea{\begin{eqnarray}}
\def\eea{\end{eqnarray}}
\def\Bphi{\mbox{\boldmath $\Phi$}}
\def\Bpi{\mbox{\boldmath $\pi$}}
\def\hphi{\mbox{\boldmath $\hat\Phi$}}
\def\Bx{\mbox{\boldmath $x$}}
\def\Bxi{\mbox{\boldmath $\xi$}}

\begin{document}
\title{The Chiral Phase Transition in Dissipative Dynamics}
\author{G. Holzwarth\thanks{%
e-mail: holzwarth@physik.uni-siegen.de} and J.Klomfass}
\address{Fachbereich Physik, Universit\"{a}t Siegen, 
D-57068 Siegen, Germany} 
\maketitle

\begin{abstract}\noindent
Numerical simulations of the chiral phase transition 
in the (3+1)dimensional $O(4)$-model are presented.
The evolutions of the chiral field follow purely dissipative
dynamics, starting from random chirally symmetric initial
configurations down to the true vacuum with spontaneously broken
symmetry. The model stabilizes topological textures which are 
formed together with domains of disoriented chiral condensate (DCC)
during the roll-down phase. The classically evolving field acts as
source for the emission of pions and $\sigma$ mesons. 
The exponents of power laws for the growth of
angular correlations and for emission rates are extracted.
Fluctuations in the abundance ratios for neutral and charged pions
are compared with those for uncorrelated sources as potential
signature for the chiral phase transition after heavy-ion collisions.
It is found that the presence of stabilizing textures (baryons and
antibaryons) prevents sufficiently rapid growth of DCC-domain size,
so observability of anomalous tails in the abundance ratios is
unlikely. However, the transient formation of growing DCC domains
causes sizable broadening of the distributions as compared to the
statistical widths of generic sources.
\end{abstract} 

\vspace{3cm}
\leftline{PACS numbers: 05.45.-a,11.10.Lm,11.30.Rd,
11.27.+d,12.39.Fe,25.75.-q,64.60.Cn,75.40.Mg}  

\leftline{Keywords: Chiral Phase Transition, Disoriented
Chiral Condensate, Sigma models,} 
\leftline{Topological textures, Skyrmions, Bags, Heavy-Ion Collisions}

\newpage
\section{Introduction}
There is general consent that the energy densities achieved in
relativistic heavy-ion collision experiments performed or planned at
RHIC or LHC should be sufficient to drive hadronic matter 
through the QCD transition into a new phase with restored chiral
symmetry and, probably, deconfined color. A commonly accepted
scenario~\cite{Bjorken} assumes that immediately after
(central) collisions most of the baryon number is still concentrated in
Lorentz-contracted pancakes receding
from the collision point with approximately the speed of light (in beam
direction), leaving behind in the c.m.
system a rapidly expanding cylindrical or spherical fireball of highly
excited "vacuum". 
Although almost void of net baryon number, the energy density
initially deposited in this 'fireball' is sufficient to
produce during the subsequent cooling phase a large number of
hadronic particles, mostly pions, but also heavier mesons and
baryon-antibaryon pairs. If inside the initial hot fireball chiral
symmetry indeed has been restored, then during the following relaxation
process the relevant order parameter, the chiral condensate, must
re-increase to finally approach the nonzero value which characterizes
the true vacuum. The time scale for this evolution of the order
parameter is set by typical relaxation times $\tau$ of the strongly
interacting system (e.g. through dissipation of energy by particle
emission) which need not necessarily be closely tied to kinematic
time scales $\tau_c$ of the collision and expansion time of the
fireball. If it is possible to replace during this relaxation process the
complex dynamics of the strongly interacting fermionic and bosonic
elementary degrees of freedom by effective dynamics for an
order-parameter field, then the time scale for changes in the relevant
effective potential can be quite distinct from the time scales of
dissipative processes which characterize the relaxation of the order
parameter towards low-energy configurations in the changing effective
potential. So, within that concept we can
study evolutions which proceed through sequences of configurations
which are far from equilibration.
Especially, for $\tau_c \ll \tau$, (the so-called "sudden quench"),
the system moves through maximally non-equilibrated sequences.

The standard tool for an effective description of hadronic physics
proceeding on energy scales of less than 1 GeV is the chiral $O(4)$
model. The isoscalar quark condensate $\left<\bar q q \right>$
(with finite value $\left<\bar q q \right>_0$ in the true vacuum)
is combined with a corresponding isovector condensate $\left<\bar q
\vec\tau q \right>$ to form a four-component "chiral" field $\Bphi$,
considered as space-time-dependent order-parameter field, and subjected
to an effective action which governs the "chiral dynamics".
(We do not discuss here the natural extension to the full SU(3) flavor
group.)
Fluctuations of $\Bphi$ describe the pseudoscalar pions and scalar 
$\sigma$-mesons. In its simplest form, the "linear $\sigma$-model",
the effective action comprises the kinetic two-gradient term and an
appropriate time- or temperature-dependent $\Phi^4$ potential. In this
form the model has been widely used for investigating features
of the chiral phase transition, both in classical dynamics~\cite{Raja},
including dissipation and noise~\cite{Chaud},
and in quantum field theoretical approaches~\cite{Boya}.

The generic structure of this model applies to
a wide variety of physical systems with spontaneously broken symmetry, 
such that particular phenomena may be expected to occur irrespective of
the nature of the underlying physical degrees of freedom.
Depending on the number of field components and spatial dimensions
the formation of ordered domains, separated by topological defects
or textures characterizes ordering evolutions in all such systems~\cite{Bray}.
This analogy has prompted the idea to use pions emitted from
differently oriented ordered domains as indicator for the chiral
phase transition itself~\cite{Anselm}. It is expected that the
temporary misalignment 
of the chiral field in finite spatial domains (as compared to the
orientation of $\Bphi$ in the true vacuum) leads to anomalous
fluctuations in the multiplicities of emitted charged or neutral pions
which may provide a signature for the transient existence of disoriented
chiral condensate (DCC).

On the other hand, for temperature $T$=0 the interpretation of
topologically nontrivial $\Bphi$-configurations 
as baryons~\cite{Skyrme} (in linear or nonlinear realizations of the
$O(4)$-symmetry) has provided surprisingly successful models
for baryon structure and dynamics. Their stabilization requires
additional terms in a gradient expansion of the effective action.
Their form is fixed by chiral symmetry, the magnitude of the few
relevant low-energy constants is extracted from experimental
information. These higher-order gradient terms incorporate at least
partly the influence of vector mesons. Naturally, the well-established
Skyrme term is the appropriate and simplest tool for this purpose.

As it is well known that stable textures play a most important role
in the ordering evolutions of multi-component fields, it appears
essential to allow for their formation and stabilization in the
dynamics of evolving field configurations. 
At the same time this opens up the possibility to describe creation and
annihilation of baryonic structures during the evolution of the fireballs.
These structures have to be supplied with a definite spatial length
scale which is provided by the stabilizing mechanism in the effective
action, which prevents them from collapse or indefinite growth. 
Using all four components of the $O(4)$-field as
independent variables implies that in the phase with restored 
symmetry the absolute value of the condensate can be very small
locally, as compared to its true vacuum value. This is different from the 
nonlinear version of the $O(4)$-model where $\Bphi$ is restricted to
the 3-sphere $\Phi^2=f_\pi^2$ and symmetry restoration can only occur
through spatial averaging over different orientations. 

The true vacuum which surrounds the fireball sets the boundary
condition for the evolution of the ordering field such that 
$\Bphi(\Bx,t)$ approaches a fixed vector for $r \to \infty$.
In that case, all configurations of the nonlinear model fall into
classes distinguished by integer winding numbers $B$. The same is true
for configurations $\Bphi(\Bx,t)$ of the linear $O$(4)-model, as long as
vanishing field vectors are excluded. While $B$ in the nonlinear model
is topologically conserved in time, in the linear version integer jumps
in $B$ may occur if at some point in spacetime $\Bphi(\Bx,t)$ moves
through $\Bphi=0$. Topologically nontrivial configurations with
nonvanishing $\Bphi$ embedded in 
the surrounding vacuum field then are interpreted as baryons with baryon
number $B$. 
In the nonlinear
version of the $O$(4)-model these 'baryons' (with baryon number $B$ and
definite size) are topologically stable against unwinding. In the
linear $O$(4)-model they represent metastable configurations which
occasionally may unwind, especially if chiral symmetry is broken
explicitly through a non-vanishing pion mass. As topological arguments
do not apply to lattice implementations, $B$-conservation if desired
has to be imposed in both cases (linear or nonlinear $O$(4)-model) by an
optional additional constraint, which rejects occasional configuration
updates that would imply a jump in $B$. 

Numerical simulations based on such very specific effective actions
which incorporate definite scales
will provide answers for physical observables which characterize
the specific physical system under consideration. So, it is not so much
our aim to look for universal features of a certain class of models,
but rather to obtain statements about specific physical situations
which hopefully comprise the essentials of a selected class of
experiments. 

If, during the initial stages of the fireball evolution, color indeed
is deconfined, then evidently, the applicability of the $O(4)$-model
is restricted to a later phase where the color-degrees of freedom are
re-frozen into the colorless order-parameter field $\Bphi$. We assume that 
this local color-confining process does not lead to any preferred local
direction of the chiral field. So, assuming that the relaxation of $\Bphi$
is driven by effective chiral dynamics, implies that color-confining
transition and the restoration of chiral symmetry may be considered
separately. 

Although dissipative evolution of the order-parameter field is a 
classical concept, the field fluctuations around the evolving
configurations acquire particle properties and must be quantized as
emitted radiation. With the classical field acting as a source for
the emission of field quanta it is expected that the radiation carries
signatures of the source field configuration. This idea underlies the
search for DCC effects in the pion yields after heavy-ion collisions.
To define the separation of the classically evolving part from the quantum
fluctuations we remove all propagating terms (second time-derivatives)
from the equations of motion for the classical configuration, i.e.
the classical evolution is defined as solution of the purely dissipative
Time-Dependent-Ginsburg-Landau (TDGL) equations. As a consequence, the
radiation is driven by the instantaneous field velocities, from which
the event-by-event fluctuations in abundance ratios may be obtained. 

The following discussion represents an extension of previous work
dealing with corresponding features in the lower-dimensional (2+1)d
$O$(3) model~\cite{O(3)}. In contrast to the exploratory nature of the
previous investigations we here attempt to define the model 
as close as possible to the scales which characterize actual
hadronic matter and to the physical situation which might prevail 
after an ion-ion collision. 
In Chpt.II we formulate the equations of motion and the resulting
emission rates. For comparison we present also the corresponding
expressions for the abundance ratios for DCC sources and for uncorrelated
stochastic sources. The model and the relevant parameters which will be
used for the numerical evolutions are specified in Chpt.III. 
In Chpt.IV the initial and boundary conditions for these evolutions
are defined, together with important observables measured during
individual evolutions, like net baryon number, baryon plus antibaryon
number, equal-time angular correlations as a measure for the size of
disoriented domains. 
Chpt.V presents numerical results, from which 
typical exponents for growth laws are extracted. These are finally used
to draw conclusions about the possibility to observe signatures of the
chiral phase transition from charge fluctuations in the pion radiation
emitted from the relaxing source.

\section{Nonequilibrium field dynamics and DCC signals}
\subsection{The TDGL equations}

We consider an effective action for an $O(4)$ vector field $\Bphi$
\be
{\cal S}[\Bphi]= \int \left({\cal T}[\Bphi]-{\cal U}[\Bphi]\right)d^3xdt
\ee
where the kinetic energy density ${\cal T}[\Bphi]$ comprises all terms
containing time-derivatives of $\Bphi$. We follow the evolution of field
configurations, which at some initial time deviate from the global
equilibrium configuration (the 'true vacuum') within some finite
spatial region $V$. The equations of motion
$\delta{\cal S}/\delta\Bphi=0$ which govern the classical evolution
of the initial non-equilibrium configuration describe both, the
approach towards the vacuum configuration in the interior of that
spatial region and the propagation of outgoing (distorted) waves into
the (initially undisturbed) surrounding vacuum. Although, of course,
the classical equations of motion conserve the total energy
$E=\int \left({\cal T}+{\cal U} \right)d^3x$, the outgoing waves carry
away energy from the interior of the spatial region in which 
the total energy $E=T+U$  initially was located.

In quantum field theory the quantization of the propagating
fluctuations describes multiparticle production of pions emitted
from the relaxing field configurations. 
In a separation
\be
\label{sep}
\Bphi(\Bx,t)=\Bphi_{cl}(\Bx,t)+\delta\Bphi(\Bx,t)
\ee
where $\Bphi_{cl}$ comprises all of the large amplitude motion
of $\Bphi$, and $\delta\Bphi$ contains only small fluctuations around
$\Bphi_{cl}$, expansion of ${\cal S}[\Bphi]$ to second order
in $\delta\Bphi$
$$
{\cal S}[\Bphi]  =  {\cal S}[\Bphi_{cl}] 
 +  \int \left(\frac{\delta{\cal S}}{\delta\Bphi(\Bx',t')}\right)
_{\left[\Bphi_{cl}\right]}
\delta\Bphi(\Bx',t')d^3x'dt' 
$$
\be
\label{exp}
 +\frac{1}{2} \int\int \delta\Bphi(\Bx',t')\left(\frac{\delta^2{\cal
S}}{\delta\Bphi(\Bx',t') 
\delta\Bphi(\Bx'',t'')}\right)_{\left[\Bphi_{cl}\right]}
\delta\Bphi(\Bx'',t'')d^3x'dt'd^3x''dt'' ~~+{\cal O}(\delta\Phi^3)
\ee
and variation with respect to $\delta\Bphi$ provides the classical
expression for the fluctuating part 
\be
\label{fluc}
\delta\Bphi(\Bx,t)=\delta\Bphi^{(0)}(\Bx,t) 
+\int {\cal G}(\Bx,t,\Bx',t')\left(\frac{\delta{\cal
S}}{\delta\Bphi(\Bx',t')}\right)_{\left[\Bphi_{cl}\right]} d^3x'dt'.
\ee
The homogeneous part $\delta\Bphi^{(0)}$, after quantization,  represents
the scattering of field quanta off the classical configuration
$\Bphi_{cl}$ (i.e. these are on-shell distorted waves).
The Green's function $ {\cal G}(\Bx,t,\Bx',t')$ of the operator
$\delta^2{\cal S}/(\delta\Bphi \delta\Bphi)|_{\left[\Bphi_{cl}\right]}$
relates emitted (i.e on-shell) radiation to the source term 
$\delta{\cal S}/\delta\Bphi|_{\left[\Bphi_{cl}\right]}$.
There is, however, no unique way
to separate the propagating fluctuations $\delta\Bphi$ from a
more or less smoothly evolving classical configuration $\Bphi_{cl}$
because both, $\Bphi_{cl}$ and $\delta\Bphi$, are integral parts of one
and the same evolving order-parameter field $\Bphi$. Conclusions drawn
from one part only, are subject to the arbitrariness of the chosen
separation.
In any case, in an equation of motion
that separately describes the evolution of $\Bphi_{cl}$
we expect a dissipative term to account for the loss of energy through
the emitted radiation:
\be
\label{deom}
\frac{1}{\tau}\dot{\Bphi}_{cl}=\left(\frac{\delta {\cal
S}}{\delta\Bphi}\right)_{\left[\Bphi_{cl}\right]}. 
\ee
Then, with eq.(\ref{fluc}), the time derivatives $\dot{\Bphi}_{cl}$
of the classical field determine the amplitudes of the fluctuations,
i.e. they serve as driving terms for the emission of field quanta.

For sufficiently small values of the relaxation constant $\tau$
the dissipative  term dominates the time evolution of $\Bphi_{cl}$, 
propagating parts get damped away and we can
replace (\ref{deom}) by the Time-Dependent Ginzburg-Landau (TDGL) equation 
\be
\label{TDGL}
\frac{1}{\tau}\dot{\Bphi}_{cl}=-\left(\frac{\delta
U}{\delta\Bphi}\right) _{\left[\Bphi_{cl}\right]} .
\ee
The potential energy functional $U$ contains no time derivatives of
$\Bphi$, therefore $\Bphi_{cl}$ has no propagating parts, i.e. it does
not pick up field momentum, and it provides a slowly moving
adiabatically evolving classical background field. 

\subsection{Particle emission and DCC signals}

For the adiabatically evolving process we consider the time $t$ as
parameter such that we have 
\be
\dot{U} 
=\int \left(\frac{\delta
U}{\delta\Bphi}\right)_{\left[\Bphi_{cl}\right]} \cdot \dot{\Bphi}_{cl} 
\: d^3x 
=-\frac{1}{\tau}\int \dot{\Bphi}_{cl} \cdot \dot{\Bphi}_{cl} \: d^3x
=-\sum_{a=1}^{4}\epsilon_a(t)
\ee
and may define (omitting the index $'cl'$)
\be
\label{epsi}
\epsilon_a(t)=\frac{1}{\tau}\int \dot{\Phi}_a(\Bx,t)^2 \: d^3x
=\frac{L^3}{\tau} \langle \dot{\Phi}_a^2(t) \rangle
\ee
as the energy carried away per time interval by particles emitted with
field velocity oriented in $a$-direction. Here "$\langle~\rangle$"
denotes the lattice 
average over a spatial $L^3$ lattice. Due to the slow motion of the
source we expect the particles to be mainly emitted with low energies,
i.e. with low momenta $k \approx 0$ and energies $\omega_a$
approximately given by their masses $\omega_a \approx m_a$. 
In our present context we choose the $O(4)$-symmetry to be
spontaneously broken in 
4-direction (selected by the surrounding true vacuum boundary condition
or by small explicit symmetry breaking, or by a small bias in the initial
configuration). Then the i=1,2,3 ('isospin') components of the
$O(4)$ chiral field constitute three 'pionic' fluctuational fields 
$\delta \Phi_i = \pi_i$ with small (explicitly symmetry-breaking) 
mass $m_{1,2,3}=m_\pi$, while the $\sigma$-fluctuation $\delta \Phi_4$
acquires a large mass $m_\sigma$ due to the spontaneously broken symmetry.
One of the three isospin directions ($i=3$, say), we identify with the
neutral $\pi_0$-mesons. So we expect the emission rates for neutral
pions ${\dot n}_0(t)$ and for charged pions ${\dot n}_{ch}(t)$ 
\be
\label{ni}
{\dot n}_0(t) =\frac{\epsilon_3(t)}{m_\pi},~~~~~~~~~~~
{\dot n}_{ch}(t) =\frac{\epsilon_1(t)+\epsilon_2(t)}{m_\pi}.
\ee

During the early stages of the ordering evolution the spatial averages
$\langle  \dot{\Phi}_i^2(t) \rangle $ will be similar for all three
isospin directions, but at late times with the formation of larger
disoriented domains they might differ appreciably. 
The relative abundance for neutral pions 
\be
\label{f0dot}
f_0[\dot{\Bphi}]=\frac{{\dot n}_0(t)}{{\dot n}_0(t)+{\dot n}_{ch}(t)}
\ee
is free of unknown constants and could serve as DCC signature, if it
is possible to separate in each individual event the small number of
late time "DCC"-pions from the background of those produced during
earlier stages, from decaying $\sigma$'s, and from other sources.

If angular ordering has been achieved within certain spatial domains
such that the angular gradients are small within
these domains then the right hand side of the TDGL-equation (\ref{TDGL})
may be dominated by a nonlinear term $\Bphi F[\Phi^2]$ which drives the 
growth of the condensate without changing its direction. During this
roll-down phase, we may approximate in the emission rates $\epsilon_i$ in
Eq.(\ref{epsi}) the velocities $ \dot{\Phi}_i(t) $ by 
$\Phi_i(t) F[\langle\Phi^2\rangle]$. In the abundance ratios (\ref{f0dot})
the function $F$ drops out, and we find in this
approximation for the contribution of one disordered domain
to the multiplicity ratio for neutral pions
\be
\label{f0}
f_0[\Bphi]=\frac{\Phi^2_3}{\sum\limits_{i=1,2,3}\Phi^2_i}
\ee
where $\Bphi$ is the classical field within that domain.
This result commonly is obtained from the coherent state formalism,
and from the foregoing we see under which conditions and limitations
it might apply to the actual dynamical process. 
In the numerical simulations we obtain the abundance ratios from
Eqs.(\ref{epsi}),(\ref{ni}) and (\ref{f0dot}). However, it is of interest
to compare the results with the consequences of the approximation
(\ref{f0}). 

From (\ref{f0}) the expected signals may be derived for the ideal case,
where $\nu$ disoriented domains of equal size have been formed.
For one single disoriented domain, with the chiral field $\Bphi$ 
uniformly aligned into some direction, the pionic field 
components ($i=1,2,3$) are parametrized as 
\be
\label{orient}
\Phi_{i=1,2,3}\propto \large(\sin\theta \cos\phi,\sin\theta
\sin\phi,\cos\theta \large) .
\ee
Then, within the approximation (\ref{f0}),
we find for the fraction of neutral pions relative to the
number of all pions emitted
\be
\label{fp1}
f(\pi_0)  = \cos^2\theta. 
\ee
In an ensemble of events
where all orientations of $\Bpi$ are equally probable
the ensemble average $\langle f \rangle$ of $f(\pi_0)$
is, of course, $\langle f \rangle=1/3$.
The probability $P(f)$ to find in one event of that ensemble the
fraction $f$ of neutral pions then is obtained from
\be
\langle f \rangle=\frac{1}{4\pi}\int f \sin\theta d\theta d\phi 
=\int_0^1 f P(f) df
\ee
as
\be
\label{Pf}
P(f)=\frac{1}{2\sqrt{f}}.
\ee

\begin{figure}[h]
\centering
\includegraphics[width=8cm,height=12cm,angle=-90]{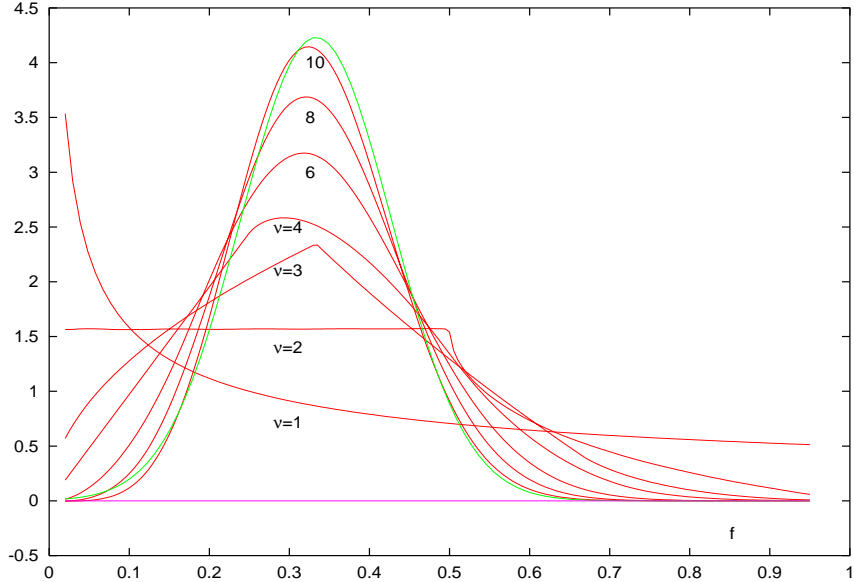}
\caption{Probability distributions $P_\nu(f)$ for
observing the fraction $f$ of neutral pions 
in an ensemble of configurations with $\nu$
disoriented domains of size $V/\nu$, according to (\ref{Pnu}).
The dashed line indicates the Gaussian (\ref{Gauss}) for $\nu=10$.}
\label{Pfig}
\end{figure}

For $\nu$ such domains of size $V/\nu$ within a source of volume $V$
the probability $P_\nu(f)$ to find the fraction $f$ in a given event
then is 
\be
\label{Pnu}
P_\nu(f)=\int_0^1...\int_0^1 \;\delta\left( f-\frac{1}{\nu}
(f_1+...+f_\nu)\right) P(f_1)...P(f_\nu)\; df_1...df_\nu
\ee
with $P(f)$ given by (\ref{Pf}). Some of these functions, for 
$\nu=1$ to $\nu=10$, are shown in Fig.\ref{Pfig}. 
For increasing $\nu$ they approach the Gaussian distribution 
\be
\label{Gauss}
P_\nu(f)=\sqrt{\frac{1}{2\pi\sigma^2_{\nu DCC}}}
\exp\left(-\frac{(f-\frac{1}{3})^2}{2\sigma^2_{\nu DCC}}\right)
\ee
with width 
\be
\label{sigdcc}
\sigma^2_{\nu DCC}=\frac{1}{\nu}\left(\langle f^2 \rangle-\langle f
\rangle^2\right) 
=\frac{1}{\nu}\left(\frac{1}{5}-\frac{1}{9}\right)=\frac{4}{45\nu},
\ee
where "$\langle \rangle$" denotes the expectation value obtained with 
the single-domain probability (\ref{Pf}).
Already for $\nu\sim 10$ domains, the distributions (\ref{Pnu}) are
very close to the Gaussian (\ref{Gauss}), with the maximum shifted only
slightly to smaller values of $f$.

For comparison, for uncorrelated sources, the probability to find the
fraction $f$ 
of neutral pions in an ensemble of events each of which emits 
(the same) total number $n$ of pions, with equal probability $p=1/3$ for
each isospin component, is given by the binomial distribution. Its
large-$n$ limit is a Gaussian 
\be
\label{Gaussbin}
P_n(f)=\sqrt{\frac{1}{2\pi\sigma^2_n}}
\exp\left(-\frac{(f-\frac{1}{3})^2}{2\sigma^2_n}\right)
\ee
with width 
\be
\label{sigbin}
\sigma^2_n=\frac{p(1-p)}{n}=\frac{2}{9n}.
\ee
(The multiplicity
distributions for neutral or charged pions commonly presented in
experimental analyses contain also the event-by-event fluctuations in
the total number $n$ of pions. For the discussion of DCC effects it
is crucial to look at the event-by-event distribution of ratios,
like $f$). The width (\ref{sigbin}) of the non-DCC distribution 
decreases with $1/\sqrt{n}$, while the width (\ref{sigdcc}) 
for $\nu$DCC events seems to be
independent of the total number $n$ of pions emitted per event.
However, both, the number $\nu=V/v_D$ of DCC domains (with fixed 
average domain volume $v_D$) present within the volume $V$,
and the total number $n$ of pions emitted from a random
source (with fixed emission density $\Gamma=n/V$), are proportional to
the total volume $V$ of the fireball. 
So, in the Gaussian limit, $\nu$ will be proportional to $n$ and we 
cannot distinguish between (\ref{Gauss}) and (\ref{Gaussbin}).
By comparison  we find $n=(5/2) \nu$, or  $v_D=5/(2\Gamma)$.
This means that a large fireball consisting of many DCC domains with
radii $R_D \sim \frac{1}{2}(5/(2\Gamma))^{1/3}$ is equivalent to a
random source with emission density $\Gamma$. For a typical value 
$\Gamma=1$ fm$^{-3}$ this radius is about $R_D\sim 0.7$ fm.
Domains growing beyond that value (for fixed $V$) lead to a broadening
of $P(f)$ according to (\ref{sigdcc}). This
broadening may occur in the Gaussian limit (especially for large
fireballs), where the number $\nu$ of DCC domains present is too large
to observe the strong anomalies in the shape of $P_\nu(f)$ for very
small numbers of $\nu$.

\section{The model}
The 3+1 dimensional $O(4)$ model is defined in terms of the
$4$-component field $\Bphi = \Phi \hphi$ with $\hphi
\cdot \hphi =1$, and the modulus field $\Phi$ of mass dimension 1,
with the following lagrangian density in $3+1$ dimensions

\be\label{lag}
{\cal{L}} = {\cal{L}}_{(2)}+{\cal{L}}_{(4)}+{\cal{L}}_{(0)}
\ee
which comprises the kinetic term of the linear $\sigma$ model 

\be
{\cal{L}}_{(2)}=\frac{1}{2} \partial_\mu \Bphi \partial^\mu \Bphi,
\ee
the four-derivative Skyrme term
\be
{\cal{L}}_{(4)}=-\frac{1}{4e^2}\left[ (\partial_\mu \hphi \partial^\mu
\hphi)^2 -(\partial_\mu \hphi \partial_\nu \hphi)(\partial^\mu \hphi
\partial^\nu \hphi) \right], 
\ee
and the potential
\be
\label{pot}
{\cal{L}}_{(0)}=- \frac{\kappa^2}{4}\left( \Bphi^2-f^2(T) \right)^2
+ H \Phi_4.  
\ee
The transition from a chiral-symmetric "hot" phase to the "cold"
phase with spontaneously broken chiral symmetry is 
driven by the coefficient $f^2$ which multiplies the quadratic term
in the potential (\ref{pot}). Generally speaking, $f^2$ is just some
input function of time (and space), which is negative in hot 
and positive in cold areas. If the transition proceeds through at least
locally equilibrated states, then $f^2$ is a function of the
time-dependent (local 
or global) temperature $T(t)$. We shall formally write $T$ for the
argument of $f^2$ and call $T$ the "temperature", although we should
keep in mind that this may be just some parametrization of the
input function $f^2$ and does not really imply that thermal
equilibration is being achieved at every point in time.

We include in the potential an explicitly symmetry-breaking term 
acting in the intrinsic 4-direction with constant 
(temperature-independent) strength $H$. 
The minimum of the potential is located at $\Bphi_{min}=(0,0,0,f_0(T))$,
related to the coefficient $f^2(T)$ of the quadratic term through
\be
\label{condens}
f^2(T)=f_0^2(T)-\frac{H}{\kappa^2 f_0(T)}.
\ee
Small fluctuations around this minimum
orthogonal to the 4-direction carry the "$\pi$-mass" $m_\pi$,
\be
\label{pimass}
m_\pi^2(T)= \frac{H}{f_0(T)} 
\ee
while the fluctuations in 4-direction are characterized by 
the "$\sigma$-mass" $m_\sigma$ 
\be
\label{sigmass}
m_\sigma^2(T)=2 \kappa^2 f_0^2(T)+m_\pi^2(T).
\ee
Conventionally, we define the "bag field" $\varphi$ by
normalizing the modulus of $\Bphi$ to its ($T$=0) vacuum value
$f_\pi=f_0(T \to 0)$
\be
\Bphi=f_\pi \varphi \hphi 
\ee
such that the bagfield $\varphi$ equals unity in the physical vacuum.

With the physical ($T$=0) values of $f_\pi$=93 MeV, $m_\pi$=138 MeV,
with the standard Skyrme parameter $e$=4.25, the only remaining free
parameter in ${\cal{L}}$ is the dimensionless coupling constant
$\kappa^2$. Through
(\ref{sigmass}) it is related to the ($T$=0) $\sigma$-mass. For the
range  6 $< m_\sigma/f_\pi <$ 10  we find a 
typical range of 18 $<$$\kappa^2$$<$ 50.
The spatial extension of a B=1 skyrmion is $(e
f_\pi)^{-1}$, so we will be dealing with baryons of typical (baryonic)
size of about 0.5 fm. (Not to be confused with their electromagnetic
radii which receive sizable contributions from vector meson clouds.)

\section{Roll-down and Domain growth after sudden quench}
\subsection{Sudden quench}

\begin{figure}[h]
\centering
\includegraphics[ width=8cm,height=12cm,angle=-90]{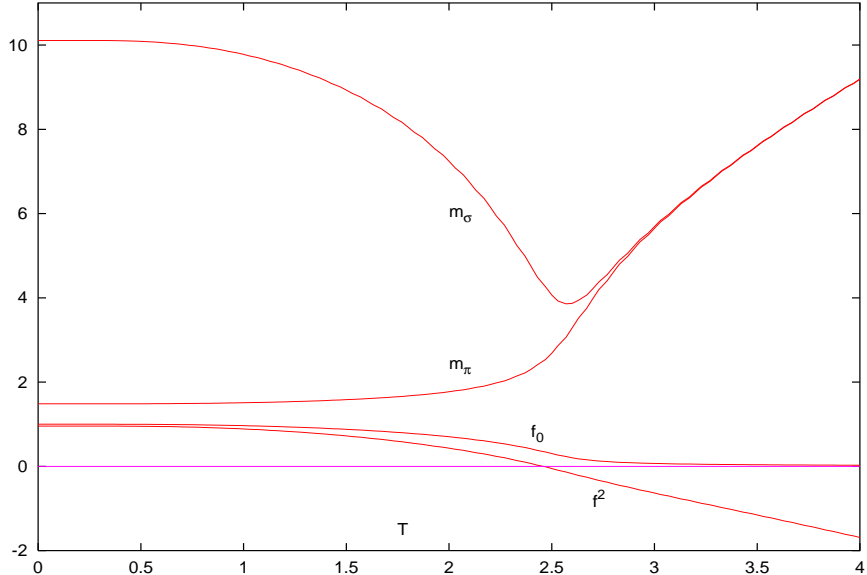}
\caption[]{ Typical features~\cite{Hans} of the condensate $f_0(T)$, 
the coefficient of the quadratic term $f^2(T)$ in the
$\Phi^4$-potential, the pion mass $m_\pi$, and the $\sigma$-mass
$m_\sigma$, as functions of the temperature $T$ for
the chiral phase transition. (Energies and temperature 
in units of $f_\pi$.)} 
\label{hans}
\end{figure}

As an input for the classical evolutions the detailed form of 
the function $f^2(T)$ in (\ref{pot}) is needed. If slow cooling proceeds
through a sequence of thermally equilibrated states the temperature
dependence of the condensate $f_0^2(T)$ may be obtained from
thermal field theory, from loops at the bosonic or the underlying quark
level. Typically, for temperatures above a critical
temperature $T_c$, the condensate $f_0(T)$ slowly approaches zero
$f_0(T/T_c >1) \to 0$. According to (\ref{condens}-\ref{sigmass}),
the $\pi$- and $\sigma$-masses then increase and come close to each other,
$m_\pi/m_\sigma \to 1$ for $T/T_c >1$, while the second-order
coefficient  $f^2(T/T_c >1)$ turns negative with increasing absolute
value (cf. Fig.\ref{hans}).
Although all of this looks very much like the standard picture
of a second-order phase transition, it probably should not be trusted
too much near and above $T_c$, because the chiral phase transition
may be closely related to the color-deconfining transition, so number
and nature of the relevant degrees of freedom may change drastically
near $T_c$, and our basic concept to describe the dynamics in terms of
a hadronic order-parameter field may break down.
 
However, if the cooling proceeds sufficiently fast as compared to the
typical relaxation time $\tau$ of the system we can impose a "sudden
quench"  
where at time $t$=0 the system is prepared in some hot $T>T_c$
initial configuration, 
and then the "temperature" is quenched instantly down to $T=0$.
Then, for times $t>0$ the system evolves from its high-$T$ initial
configuration, subject to the $T$=0 form of the potential~(\ref{pot}).
Clearly, in this case no assumption about thermal equilibration, 
even more, no further information on the specific form of $f_0(T)$
is needed.

\subsection{Initial and boundary conditions}
Evolution  of $\Bphi_{cl}$  proceeds through the
TDGL equations of motion as they result from the potential part of
(\ref{lag}), (i.e. also from the
Skyrme term only the spatial-derivative terms are kept).
As far as the dissipative term originates from elimination of
fluctuational modes which simulate a heat bath for the evolving field
configuration, addition of fluctuational noise $\Bxi$ may be required.

With the equations of motion (\ref{TDGL}) being purely first order in time
derivatives it is sufficient to 
specify for initial conditions at  $t=0$ only the field configurations
themselves. As time-dependent dynamic fluctuations are not part of
$\Bphi_{cl}$ the high-temperature initial configuration is
$\Bphi_{cl}\equiv 0$ in the interior of the spatial region
in which the hot chiral field is located
(apart from a small optional bias in 4-direction due to explicit symmetry
breaking), and it takes on the true vacuum values on its boundary. The
first step in the time evolution 
is therefore governed by the stochastic force $\Bxi$ alone. But this
first step will immediately create a configuration $\Bphi_{cl}$ which is of
stochastic nature itself. Therefore, choosing 
each of the four cartesian field components at each interior point
of the lattice  from a random Gaussian deviate, and fixing $\Bphi=
(0,0,0,f_\pi)$  on the boundary of the lattice, provides convenient initial
configurations with spatial correlation length less than one
lattice unit. The width $\Delta$ of the Gaussian deviate reflects the 
initial temperature $T>T_c$. It is not really necessary to specify a definite
value for this temperature because after about one relaxation-time unit
(after the sudden quench) the details of the initial configuration are
lost anyway and evolutions proceed quite similarly, for widths
chosen in the range $0.1<\Delta/f_\pi<0.5$. Numerically, very small
values of $\Phi$ (i.e. small values of the width $\Delta$) 
require very small timesteps at the beginning, therefore
for convenience we generally choose $\Delta/f_\pi =0.3$.
The $(T=0)$ vacuum boundary conditions 
\be
\label{bc}
\Bphi(t)=(0,0,0,f_\pi)
\ee 
are kept fixed for all times at the lattice surface. This allows
compactification of 3-space to a 3-sphere $S^3$, and guarantees integer
winding numbers.

Starting from these initial conditions events are generated by evolving
the configurations according to the TDGL eq.(\ref{TDGL}). Because the
time scale enters only through the $\dot{\Bphi}$ term, it is convenient
to measure the time in relaxation-time units $\tau$.
A large number of events generated in this way constitute the
statistical ensemble from which ensemble averages then can be obtained at
some point in time during the evolution. 

\subsection{Conservation of baryon number}

The topological current is
\be\label{top}
j_\mu =\frac{1}{12 \pi^2} \epsilon_{\mu \nu \rho \sigma}
\epsilon_{a b c d} \hat\Phi_a \partial^\nu\hat\Phi_b
\partial^\rho\hat\Phi_c\partial^\sigma\hat\Phi_d
\ee
which satisfies $\partial^\mu j_\mu = 0$. 
It allows to assign a value $\rho(i,j,k)$ of the 
winding density $\rho \equiv j_0$ to each point $(i,j,k)$ of the cubic
lattice (or, more precisely, to each elementary lattice cube 
with lowerleft corner $(i,j,k)$)
such that the total winding number
\be
B=\sum_{i,j,k=0}^{L-1} \rho(i,j,k)
\ee
summed over the whole lattice is integer and configurations can
be selected with some desired integer value of $B$. We also define 
\be
\label{N}
N=\sum_{i,j,k=0}^{L-1} |\rho(i,j,k)|
\ee
by summing up the absolute values of the local winding densities.
Of course, $N$ generally is not an integer. However, if a configuration
describes a distribution of localized textures (and antitextures) which
are sufficiently well separated 
from each other, then $N$ is close to an integer and counts the number
of these textures (plus antitextures). In that case we can define the
numbers $N_+,N_-$ 
of "baryons" and "antibaryons" through
\be
B=N_+ - N_-~~~~~~~~~~~N=N_++N_-~~.
\ee

Even if it is not integer we will in the following sometimes briefly
call $N$ the "number of baryons plus antibaryons". 

In a lattice implementation the local update of $\Bphi(i,j,k)$ at some
timestep at some lattice point $(i,j,k)$ will occasionally lead to a
discrete jump in the total winding number $B$. 
For well-developed
localized structures this corresponds to unwinding textures or
antitextures independently, such that $B$ decreases or increases by one
or more units.
This will eventually happen even for implementations of the nonlinear
$O(4)$-model where the length $\Phi$ is constrained to $f_\pi$ and $B$ is
topologically conserved, because the topological arguments based on
continuity do not apply to the discrete lattice configurations. 
It can be prevented by an (optional) $B$ filter which rejects such a
$B$-violating local field update at that lattice point and timestep.
This eliminates all independent unwinding processes. Only 
simultaneous annihilation of texture and antitexture in the same time
step remains possible, and, as the update proceeds locally at each
lattice vertex it can happen only if texture and antitexture overlap.
This $B$-conserving evolution is characteristic for the nontrivial
topology of the nonlinear $O(4)$-model and in this way can be
implemented as an optional constraint also into the linear $O(4)$-model. 
In the continuum limit it implies that a vanishing modulus $\Phi=0$ of the
field vector is excluded.

Practically, during the very early stages of an evolution, where the
local gradients of the angular fields still are of the order of $\pi/a$,
(where $a$ denotes the lattice
constant), annihilation processes by far exceed local unwinding, and cause
a very rapid decrease of the initially large number $N$. 
During that stage the definition of the global baryon number $B$ 
requires a prescription of how to define the baryon number located on
one elementary lattice cube (which necessarily involves some
arbitrariness like mapping on geodesics). 
If the lattice
constant $a$ is chosen sufficiently small (as compared to the typical 
spatial size $R=(e f_\pi)^{-1}= 0.5$ fm of stable $B=1$ baryons) 
such that the emerging localized extended structures are described with
reasonable accuracy on the lattice, the small global 
baryon number $B$ stabilizes very quickly and local unwinding no longer
occurs. It turns out that $R/a=3$ is
sufficient to avoid the need for imposing an explicit constraint on
$B$. This implies a lattice constant of $a = 1/6$ fm, which appears
as a reasonable value for the lattice implementation of an effective
low-energy model. 

As there is no universal simple scaling law (due to the ocurrence 
of different powers of gradients in ${\cal{L}}$, or in other words, 
due to explicit scales introduced through soliton size and symmetry
breaking) changes of the lattice constant can only partly be absorbed 
into corresponding changes in the time scale, and we have to check to
which extent physical statements, like the physical size of oriented
domains at the time when the roll-down is completed, are independent of
the choice of the lattice constant $a$ 
(for a fixed physical size $(aL)^3$ of the finite total volume).

\subsection{Angular correlations}

A convenient measure for the average size $R_D$ of (dis)oriented
domains is the half-maximum distance $R_{1/2}$
of the equal-time lattice averaged angular correlation function 

\be
\label{corr}
C ( R ) =\left[
 \underbrace{\sum_{i, j, k = 0}^{L} \ \sum_{l, m, n = 0}^{L}}_{
\delta\leq R < \delta+1}
  \hphi(i, j, k) \cdot \hphi ( l, m, n)
\;/ \underbrace{\sum_{i, j, k = 0}^{L} \  \sum_{l, m, n = 0}^{L}}_{
\delta\leq R < \delta+1}  \ 1 \right]  - \langle \hphi \rangle^2
\ee
\begin{displaymath}
{\rm with} \qquad \delta=\sqrt{(i-l )^2 + (j-m )^2 + (k-n )^2}                        
\end{displaymath}

Note that the definition (\ref{corr})
contains only the unit vectors $\hphi$, and not the full field
vectors $\Bphi$, i.e. $C(R)$ measures only the {\it angular}
correlations in a given configuration. 
The half-maximum distance $R_{1/2}$ is conveniently
obtained by interpolating between the two neighbouring integer values of
$R$ where $C(R)$ passes through $C(0)/2$. This lattice averaged correlation
length can be obtained as function of time for any individual
configuration (in a statistical ensemble of many events) and compared to
the domain pattern of that same individual configuration. Analysing the
domain sizes shows that $R_D \approx R_{1/2}$ provides a reasonable
measure for the average size $R_D$ of (dis)oriented
domains. To extract an accurate growth law for $R_D$, of course, would
require to perform an additional ensemble average at each point in
time, but for a sufficiently large lattice the spread in $R_{1/2}(t)$ 
for different events is small (as long as $R_{1/2} < L/2$), and it is
sufficient to study $C(R)$ on an event-by-event basis.

Explicit symmetry breaking (through nonvanishing $H$ in
(\ref{pot}), and through the vacuum boundary conditions (\ref{bc})
imposed on $\Bphi$) causes a nonzero average $\langle \hphi
\rangle$  growing (in 4-direction)  with time, 
so on a finite lattice at late times
it becomes numerically inconvenient to extract $R_{1/2}$ from $C(R)$. 
Therefore, in analogy to (\ref{corr}), we define a correlation function
$C_\pi(R)$
\be
\label{corrpi}
C_\pi ( R ) =
 \underbrace{\sum_{i, j, k = 0}^{L} \ \sum_{l, m, n = 0}^{L}}_{
\delta\leq R < \delta+1}
  \hphi_\pi(i, j, k) \cdot \hphi_\pi ( l, m, n)
/ \underbrace{\sum_{i, j, k = 0}^{L} \  \sum_{l, m, n = 0}^{L}}_{
\delta\leq R < \delta+1}  \ 1
\ee
where $\hphi_\pi$ is the "pionic" (orthogonal to the 4-direction) 
part of $\hphi$, renormalized to form a 3-dimensional unit vector
$\hphi_\pi\cdot\hphi_\pi =1$. The function $C_\pi(R)$ 
measures the angular correlations in the pionic 3-space
orthogonal to the 4-direction, unaffected by explicit symmetry
breaking. It satisfies $C_\pi(0)=1$, and the corresponding half-maximum
radius will be denoted as $R_\pi$. 

The angular ordering is driven by the terms ${\cal{L}}_{(2)}$ and
${\cal{L}}_{(4)}$, which contain two, respectively four, angular
gradients. From simple scaling arguments we expect each of these
terms, separately, to lead to $t^{1/2}$, respectively $t^{1/4}$,
power-laws for the 
growth of the spatial scale of disoriented domains. However, their
relative strength introduces a definite scale into the system (the size
$(e f_\pi)^{-1}$ of the baryons), and the presence of extended textures
will cause deviations from the scaling growth laws. 

\section{Results}
\subsection{Roll-down times}

As a global measure for the restoration of the chiral condensate we 
consider the lattice average of the modulus of the order parameter
field $\langle \varphi \rangle$. In the interior of stable baryons
the $\varphi(\Bx,t)$-field will stay small (constituting the bags),
while in the exterior region it will evolve towards the vacuum value
$\varphi$=1. 
Depending on the size $R^3$ and number $N$ of baryons (+ antibaryons) 
which finally stabilize inside the total volume $(aL)^3$, the average
$\langle \varphi \rangle$ 
will approach a value near $1-N(R/aL)^3$. 
We define the typical roll-down time
$t_{RD}$ as that point in time after the sudden quench 
when $\langle \varphi \rangle$ has reached 80\% of its final value.
Locally, the roll-down towards
the vacuum value  $\varphi$=1 in the exterior regions is driven by the
potential part ${\cal{L}}_{(0)}$~(\ref{pot}) of ${\cal{L}}$. 
Therefore, for a large
lattice, as long as the baryon(+antibaryon) density is small, we expect
$t_{RD}$ to scale with $\kappa^{-2}$, i.e. with the inverse strength of the
$\Phi^4$-potential. The typical range for the ($T$=0) $\sigma$-mass 
( $6< m_\sigma/f_\pi < 10$ ) provides limits for $\kappa^2$  
($18 < \kappa^2 < 50$). This constitutes a relatively
weak potential to compete with the gradient terms in
${\cal{L}}$, so we expect deviations from a pure
$\kappa^{-2}$-power law.
Similarly, the actual roll-down times are significantly
reduced through the symmetry-breaking imposed by the finite pion mass
and the vacuum boundary conditions which become increasingly important 
for smaller lattices.   We find typical roll-down times 
of the order of $t_{RD}/\tau \approx$ 2-3 (for $\kappa^2 = 50$ and lattice
sizes $L=30-60$) 
and $t_{RD}/\tau \approx$ 5-8 for ($\kappa^2 = 20$).
This shows that the potential part indeed has a dominating influence
for the evolution of the chiral condensate. The slight prolongation of
$t_{RD}$ with increasing lattice size $L$ reflects the influence of the
vacuum boundary condition on the growth of $\Phi$ in the interior of
the volume.

\subsection{Features of the evolutions}

Evolutions after a sudden quench show three typical distinguishable phases:

\noindent I.) During the first phase, which extends to about $t/\tau
\sim 0.1-0.2$ after the quench, the initially random
configuration quickly aligns locally, such that the angular correlation
lengths $R_D \approx R_\pi$ grow to about 3-4 lattice units
(corresponding to about 0.5-0.7 fm). 
\begin{figure}[h]
\begin{center}
\includegraphics[ width=7.5cm,height=7.5cm,angle=-90]{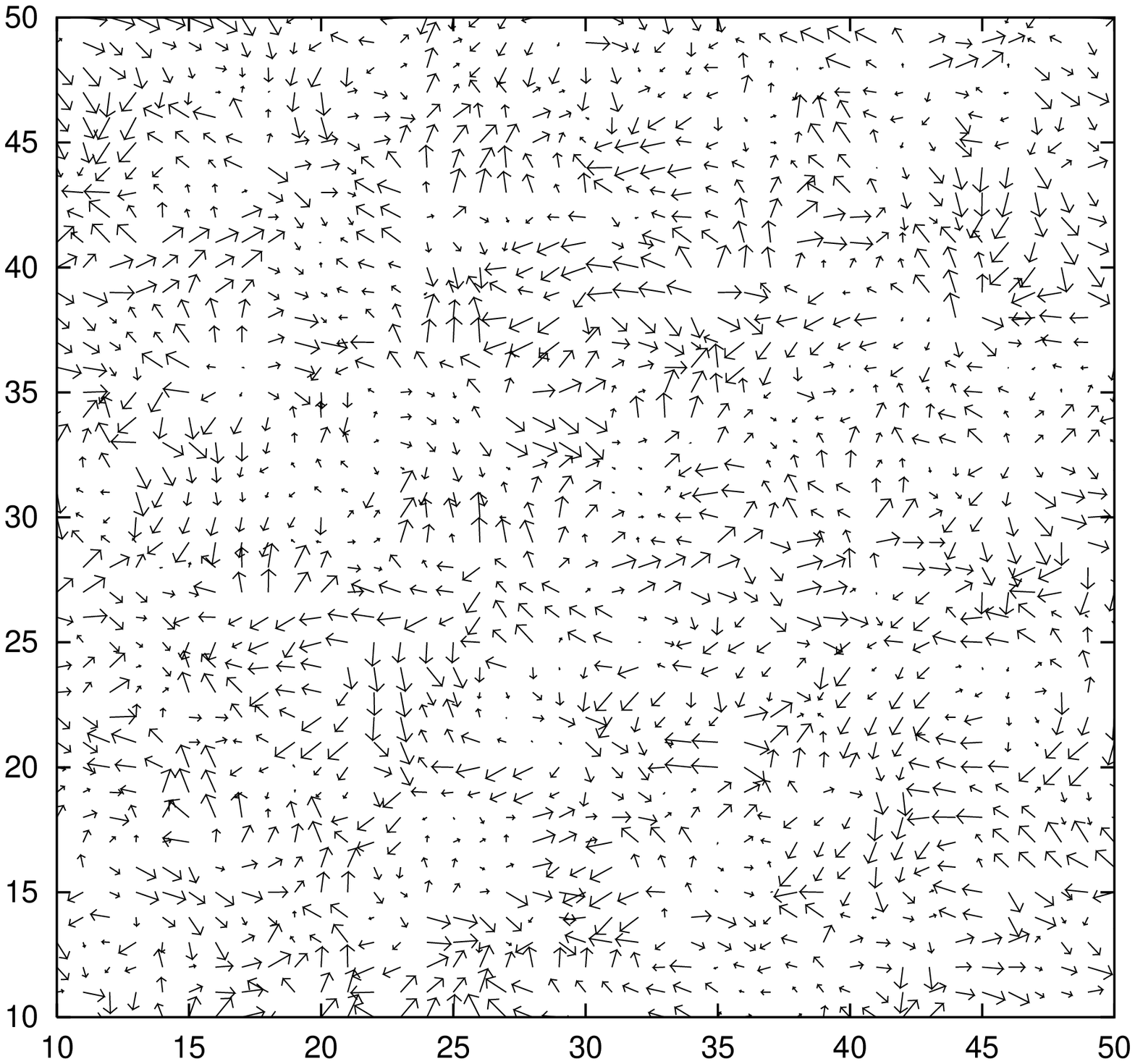}
\includegraphics[ width=7.5cm,height=7.5cm,angle=-90]{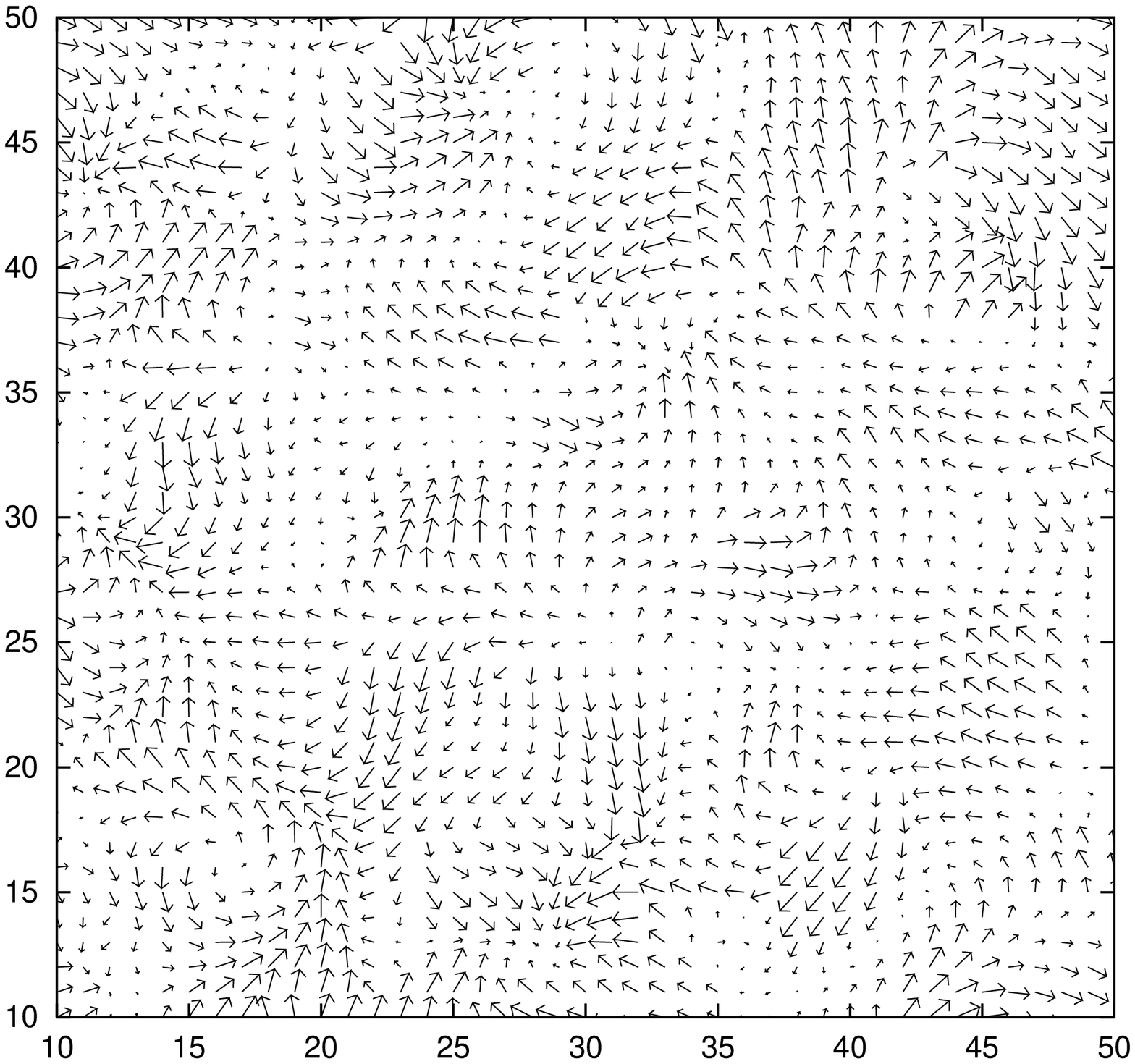}
\end{center}
\caption{ \label{field} Typical field configurations during the early phase
(I), at times $t/\tau$=0.02 and $t/\tau$=0.1 after the sudden quench.
The arrows show the isospin (unit-)vectors at the lattice points of
a 2-dimensional cut through a 60$^3$ cubic lattice, projected on a
plane in isospace. }
\end{figure}
Fig.(\ref{field}) shows a 2-dimensional cut through a 60$^3$ cubic
lattice with the isospin (unit)-field vectors projected into a plane in
isospace, at times $t/\tau=$ 0.02 (where local aligning is already
clearly visible), 
and at $t/\tau=$ 0.1, which approximately marks the end of this first
phase. The condensate $\langle \varphi \rangle$
does not grow during this period, in fact, the lattice average
decreases to a minimum slightly below its small initial value. The
total baryon-plus-antibaryon number $N$ (which is becoming really
well-defined only towards 
the end of these very early stages) drops to values of about 100-200 on
a $L$=60 lattice, corresponding to a (baryon+antibaryon) 
density of 0.1-0.2 fm$^{-3}$,
while the integer baryon number $B$ is close to or equal to zero. Already 
by the end of this short first phase the winding density has developed
pronounced maxima and minima at the locations of the emerging
(anti-)baryons, while the bag field $\varphi$ drops smoothly from its
vacuum boundary condition to 
its small value throughout the interior volume, and develops 
dips near the extrema of the winding density.
All of these features are basically
independent of the $\Phi^4$-potential strength $\kappa^2$, the total
volume $L^3$, and the symmetry breaking. The latter, however, causes a
steady increase in the lattice average $\langle \Phi_4 \rangle$
of the 4-component of the vector field $\Bphi$, which initially is 
very close to zero. 
The energy density $e=U/(aL)^3$ in the random configuration at time $t$=0
is extremely high (of the order of 10-50 GeV/fm$^3$), and located almost
completely in the angular gradients of the Skyrme term ${\cal
L}^{(4)}$. (Note that the energy density $e_V=(\kappa^2/4)f_\pi^4 $ 
located in the $\Phi^4$-potential for $\Phi\equiv 0$ is only
about $e_V$=0.05 GeV/fm$^3$ for $\kappa^2=20$.)
By the end of this first phase the energy density has dropped to values
of about 0.3-0.5 GeV/fm$^3$, i.e. to values where we might hope that
this effective 
hadronic model may become reasonable to apply. So we may conclude
that this first phase of the evolutions rather serves to create 
an ensemble of initial conditions for the onset of the physically reasonable
time-development of the model which begins near $t/\tau \approx 0.1$ after
the sudden quench with correlation lengths of the order of 0.5 fm.
\begin{figure}[h]
\centering
\includegraphics[ width=8cm,height=12cm,angle=-90]{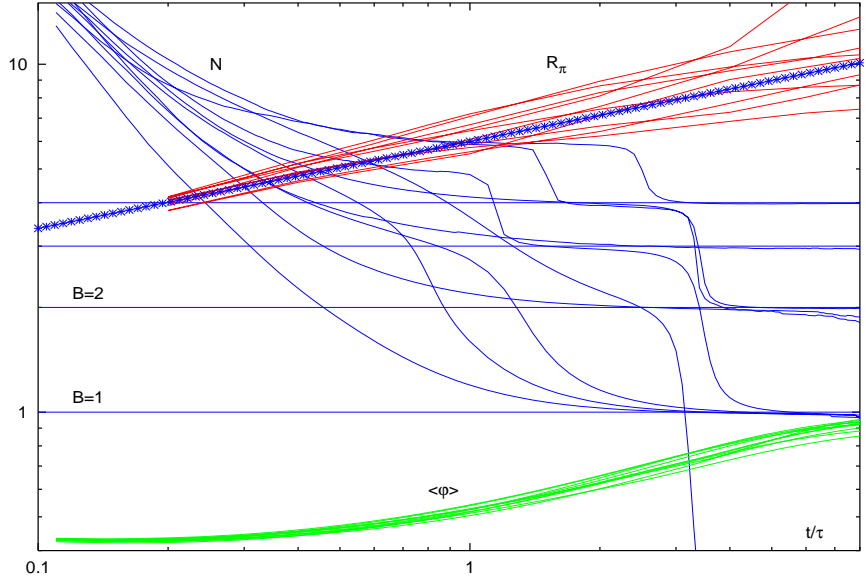}
\caption{\label{rd30} Particle numbers $N$, winding numbers $B$, 
correlation lengths $R_\pi$ (in lattice units), and average 
condensates~$\langle\varphi\rangle$, for a sample of ten events during
the roll-down phase (II) after a sudden quench. For comparison, the
straight line (with crosses) indicates the power law $6 t^{0.25}$).
The evolutions proceed on a $L=30$ cubic lattice, with potential
strength $\kappa^2=20$.}
\end{figure}

II.) The second ('roll-down') phase is characterized by the growth of
the local $\varphi$-field towards unity in the spatial regions between
the emerging baryons, so the lattice averaged condensate 
$\langle \varphi \rangle $ increases towards its final value. 
As discussed above, the duration of this roll-down phase is
dominated by the strength of the $\Phi^4$ potential and by symmetry breaking. 
Fig.(\ref{rd30}) shows a sample of ten events
evolving on an $L=30$ lattice with lattice constant 
$a=(1/3)(e f_\pi)^{-1}=0.167$ fm, $\kappa^2=20$, and $m_\pi=138$
MeV, with vacuum boundary conditions after a sudden quench.
By the time of $t/\tau=0.1$ the total 'particle' numbers have dropped to
$N \sim$ 15-20 and keep falling rapidly until after some final annihilations
they approach the fixed (net baryon) winding numbers $B$, which scatter
between 0 and 4 in this sample. The growth of the average condensate
$\langle \varphi \rangle$ is very similar for all events and differs only
through slightly varying final values due to the different baryon numbers.
\begin{figure}[h]
\begin{center}
\includegraphics[scale=0.93]{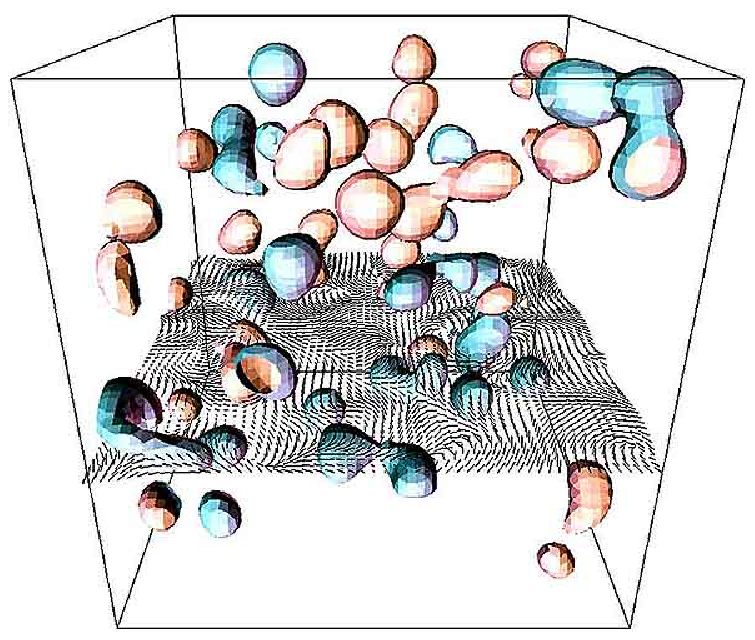}
\includegraphics[scale=0.93]{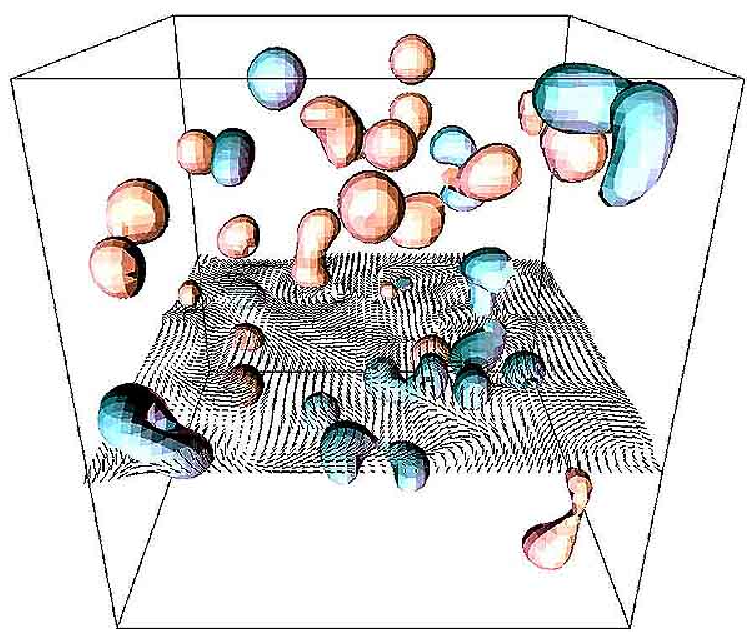}
\end{center}
\begin{center}
\includegraphics[scale=0.93]{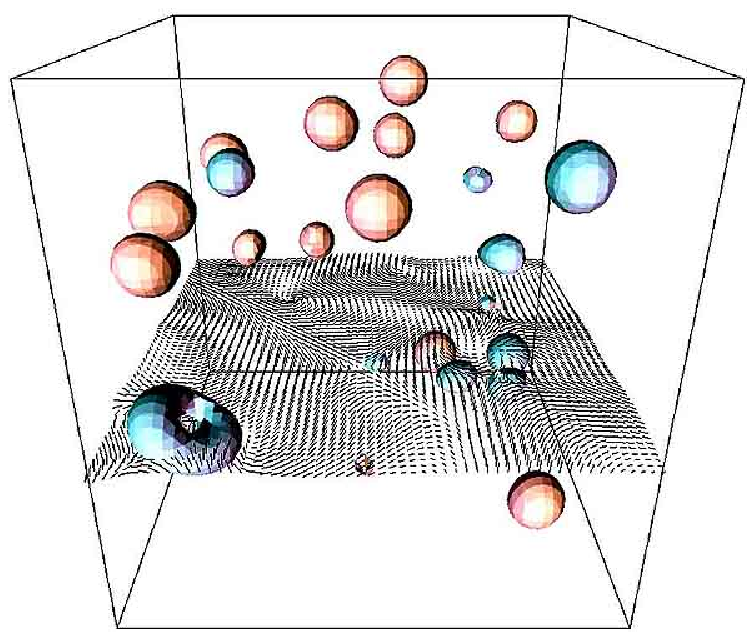}
\includegraphics[scale=0.93]{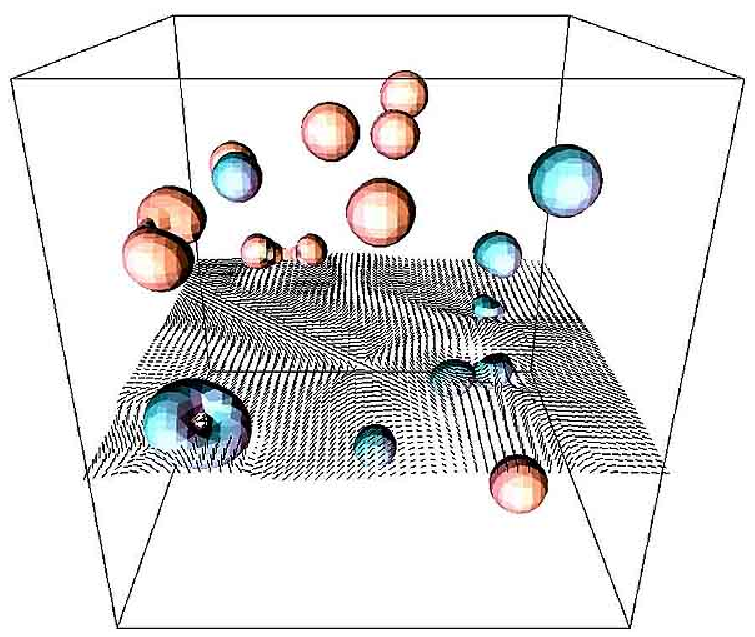}
\end{center}
\begin{center}
\includegraphics[scale=0.93]{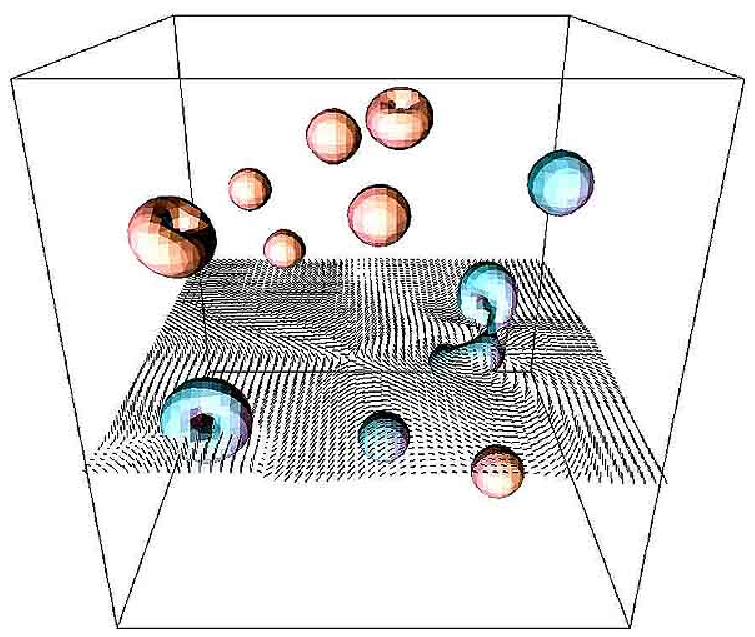}
\includegraphics[scale=0.93]{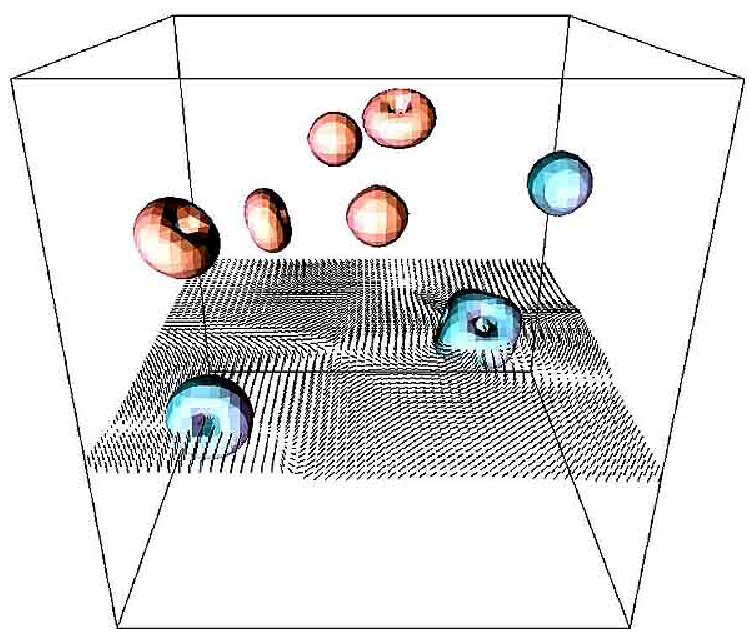}
\end{center}
\caption{ \label{plots} Snapshots of surfaces of equal (absolute)
values of winding densities during and beyond the roll-down phase
($t/\tau$=0.4, 0.8, 4.0, 8.0, 20, 40 from upper left to lower right) 
for a single event evolving after a sudden quench on a $L$=60 
cubic lattice with $\kappa^2$=20. In an arbitrarily selected
2-dimensional cut through the lattice the isospin (unit-)vectors are 
also plotted to show the growth of aligned domains.} 
\end{figure}

The angular ordering increases by moving the boundaries of oriented
domains, by further annihilations, or fusion
of individual baryons into more complex multi-$B$ configurations.
As long as such extended localized textures are present, they prevent
straight alignment of the isovector part of the field in the spatial domains
which surround and separate them, while the 
4-components follow the driving symmetry breaker towards the aligned
vacuum. Therefore, the growth of the angular correlation length $R_\pi$
is quite slow. During the roll-down phase it follows approximately a
power growth law $t^\alpha$ with $\alpha$ varying between 0.25 and 0.3, 
as we might expect from the separate or at least dominant acting of the
Skyrme term. Due to this slow growth, the average radius $R_\pi$ of
aligned disoriented domains in isospace reaches only about 7 to 10
lattice units at roll-down time $t_{RD}$, and differences in $t_{RD}$
due to different choices of $\kappa^2$ do not lead to sizeable
differences in $R_\pi$ by the end of this roll-down period. Depending
on the baryon number remaining in the configurations after roll-down,
the angular pionic correlation lengths $R_\pi$ either saturate near 10
lattice units, or (for $B=0$) keep rising towards lattice size.
However, by that time, the topologically trivial domains of the whole
field are already oriented in 4-direction, the condensate $\langle
\Phi\rangle$ 
is saturated by $\langle \Phi_4 \rangle$, i.e., the pionic
components $\Phi_i$ for $i$=1,2,3 in the aligned domains practically
vanish. In other
words, outside of stable textures the field is aligned in vacuum
direction before the (T=0) condensate is fully restored. The same holds
also for the stronger potential $\kappa^2=50$ (i.e. $m_\sigma/f_\pi
\sim 10$).

III.)  The final part of the evolution after completion of the
roll-down is characterized by the stable textures approaching their
minimal-energy configurations. Occasionally, some final annihilations
may occur, and due to mutual attraction the locally separated baryons
drift towards each other and finally combine into the multi-$B$
'nuclei' which constitute the classical minimal-energy configurations
of the model lagrangian (\ref{lag}). On the relaxation-time scale this
is an extremely slow process which extends over $\sim 10^3$ relaxation
time units,
with correspondingly small field velocities. So, from the point of
pion emission abundances, it is uninteresting. Furthermore, with
propagating terms included in the classical equations of motion, the
individual baryons or antibaryons would be able to leave the volume
before fusing into multi-$B$ configurations.

In order to convey a more pictorial view of the field evolutions we plot
in figs.(\ref{plots}) 3-d views of the winding densities $\rho_0$ from 
eq.(\ref{top}) at six different times $t/\tau$=0.4, 0.8, 4, 8, 20, 40,
which cover the roll-down period for an individual evolution on a 60$^3$
lattice. The surfaces plotted show the 
equal-(positive or negative)-density surfaces, so baryons and
antibaryons can be 
distinguished. Each snapshot contains also an arbitrarily selected plane
in which the isospin field (unit-)vectors are plotted as short lines,
so the growth of aligned domains is clearly visible. Evidently,
already at the beginning of the roll-down, the individual
(anti-)baryons are separately developed, and while the average condensate is
growing towards its vacuum value, the angular ordering proceeds mainly
through annihilation and fusion, where in this specific example several
$B=2$ torus configurations and one $B=4$ cluster are created.

\begin{figure}[h]
\centering
\includegraphics[ width=8cm,height=12cm,angle=-90]{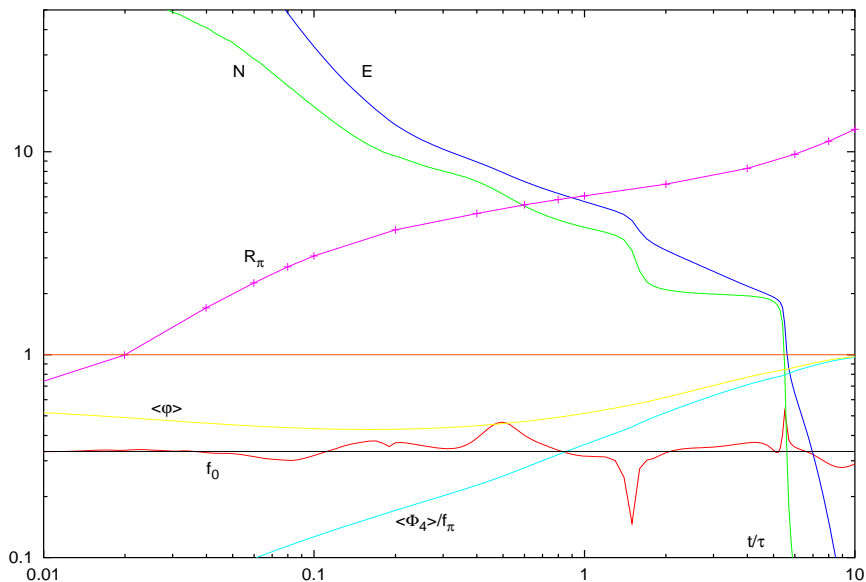}
\caption{\label{A7} Particle number $N$, total energy $E$ (in units of
$6\pi^2f_\pi/e$), correlation length $R_\pi$ (in lattice units), 
average condensate $\langle\varphi\rangle$, 
average 4-component $\langle\Phi_4/f_\pi\rangle$, and the abundance
ratio $f_0$ for neutral pion emission, obtained from the field
velocities according to eq.(\ref{epsi}-\ref{f0dot}), for one single
event during the roll-down phase (II). 
The evolution proceeds on an $L=30$ cubic lattice after a sudden quench,
with potential strength $\kappa^2=20$.}
\end{figure}

\subsection{Pion abundance ratios}

Within our approximation scheme, the abundance of neutral pions emitted
relative to all pions is given by the ratio of the averaged square of
the field velocities (\ref{epsi}-\ref{f0dot}).  
In Fig.(\ref{A7}) we look at one typical single event
evolving on a $L=30$ lattice like the sample shown in Fig.(\ref{rd30}). 
It is characterized by baryon number $B=0$. The baryon-plus-antibaryon
number approaches the integer value $N$=2 near $t/\tau\sim 1$, the final
annihilation process takes place between $t/\tau$=5 and $t/\tau$=6. By
that time 
the average condensate $\langle \varphi\rangle$ has reached 80\% of its
final value, while the average $\langle \Phi_4/f_\pi \rangle$
has almost caught up with $\langle \varphi \rangle$, i.e. even before the
roll-down is finished, the field surrounding the 
last two remaining baryons is essentially oriented in 4-direction. 
During the roll-down phase from $t/\tau\approx 0.2$ until
$t/\tau\approx 6$ 
the correlation length $R_\pi$ grows according to 6$t^{0.25}$ and
reaches about 10 lattice units at roll-down time. Only after
the final annihilation the growth rate accelerates.

\begin{figure}[h]
\centering
\includegraphics[ width=7.cm,height=12cm,angle=-90]{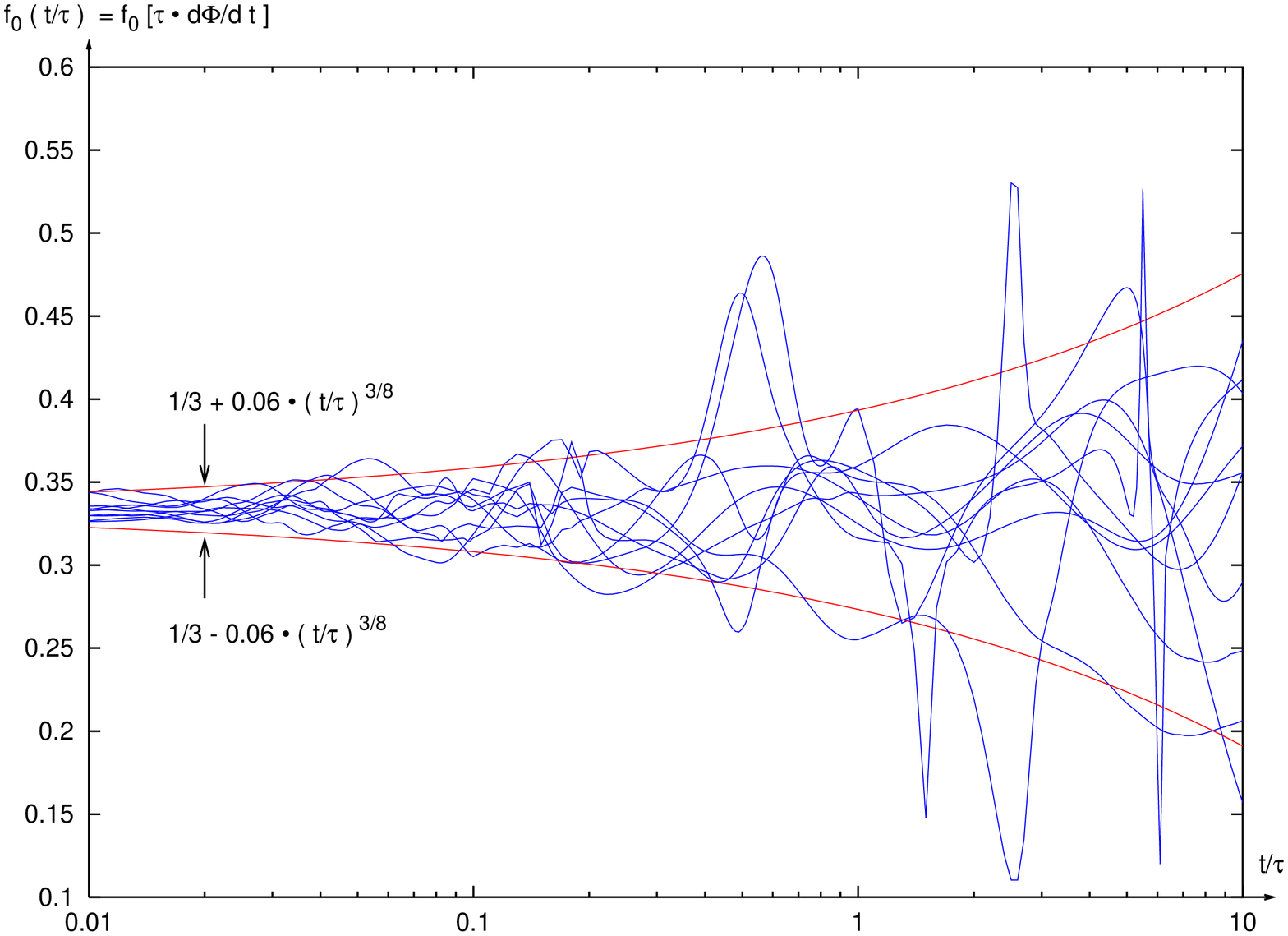}
\includegraphics[ width=7.cm,height=12cm,angle=-90]{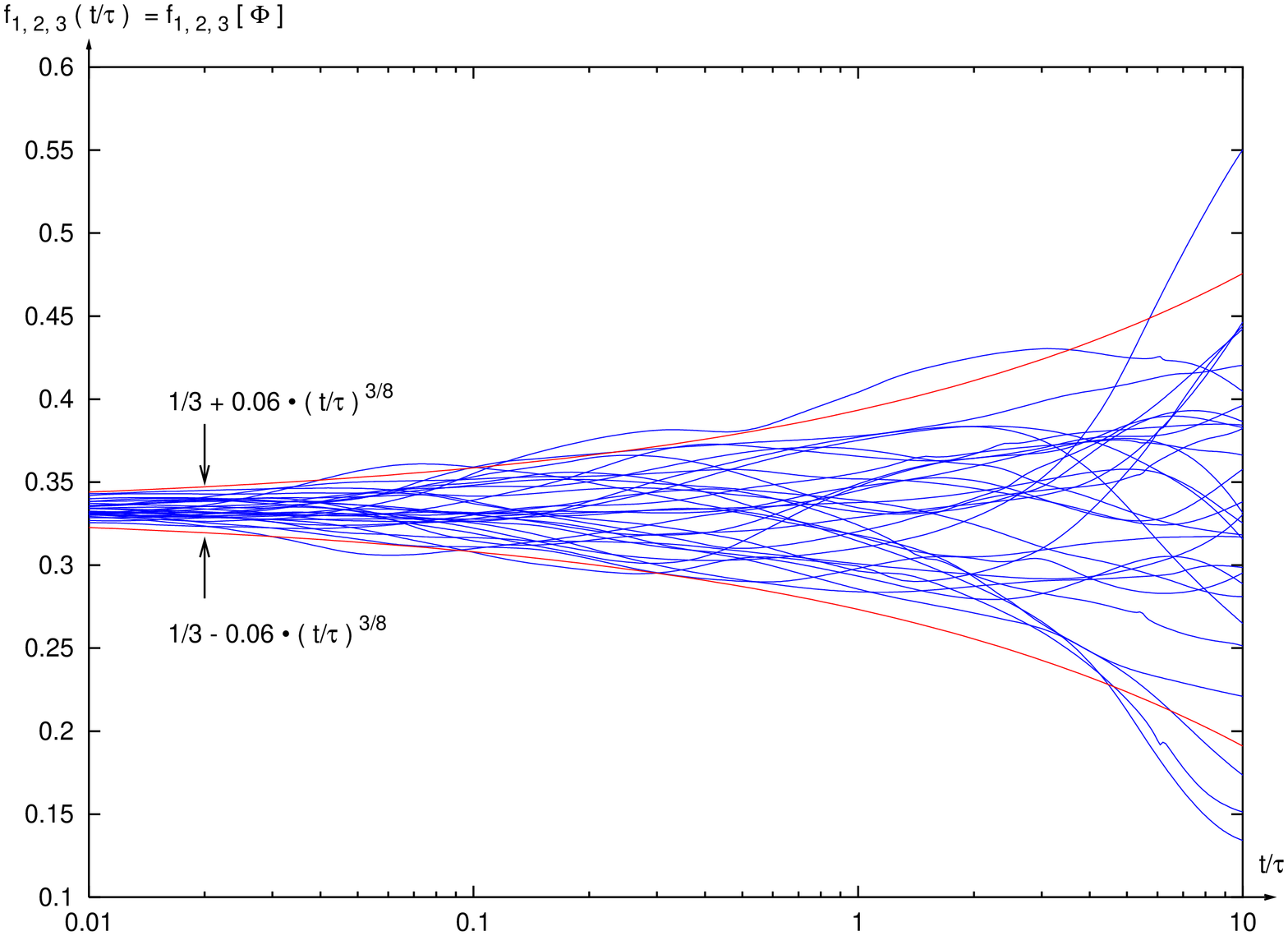}
\caption{\label{width}
a) The abundance ratios $f_0[\dot{\Phi]}$ for neutral pion emission, 
obtained from the field velocities according to
eqs.(\ref{epsi}-\ref{f0dot}) for a sample of 10 events evolving on an
$L=30$ lattice with $\kappa^2=20$. The lower part
b) shows for the same sample the ratios $f_i[\Phi]$ for all three
isovector components $i=1,2,3$, obtained from the field configurations
according to eq.(\ref{avfi}). 
For comparison the growth law $1/3\pm 0.06 (t/\tau)^{3\alpha/2}$ 
for $\alpha=0.25$ is included.}
\end{figure}

Figure (\ref{A7}) includes the abundance ratio $f_0[\dot\Phi]$ for
neutral pion emission, obtained from the field velocities as given in
eqs.(\ref{epsi}) and (\ref{f0dot}). The velocity-dependent $f_0[\dot\Phi]$
shows a few pronounced extrema, which are clearly correlated with 
major reordering in the topological structures, i.e. in this case
$B-\bar{B}$ annihilations. Otherwise $f_0$ shows small fluctuations
around the average value of 1/3.
We expect that for anomalies in pionic abundance ratios the sizes of
aligned domains with different orientations of their isospin (1,2,3)-field
components are important.
Apart from the specific fluctuations in the velocities which occur in
connection with annihilation or fusion processes of the emerging
baryons, the pionic abundance ratios calculated from averaging the
square of the field velocity components (\ref{f0dot}), or by averaging the
square of the field components  
\be
\label{avfi}
f_i[\Phi]=\frac{\langle \Phi^2_i \rangle}
{\sum\limits_{j=1,2,3}\langle\Phi^2_j\rangle}.
\ee
show comparable deviations from
the average value of 1/3. According to (\ref{sigdcc}) the widths
$\sigma_{\nu DCC}$ of these latter deviations (\ref{avfi}) grow with
$1/\sqrt{\nu}$. The number 
$\nu$ of aligned disoriented domains decreases as $R_\pi^{-3}\propto
(t^\alpha)^{-3}$, i.e. $\sigma_{\nu DCC}$ grows like $ \sim
t^{3\alpha/2}$. (In this 
consideration we neglect that the effective volume available for
disoriented domains is only the difference between the total volume and
that part which is already fully aligned in 4-direction).
In Figs.(\ref{width}a,b) we compare the ratios $f_0[\dot\Phi]$ and 
$f_i[\Phi]$ for a sample of events. Evidently, apart from the sporadic
peaks in $f_0[\dot\Phi]$ due to $B-{\bar B}$ annihilations both 
distributions are reasonably well bound by the $t^{3\alpha/2}$ law which
links them to the growing size of disoriented domains.

\begin{figure}[h]
\centering
\includegraphics[ width=8cm,height=12cm,angle=-90]{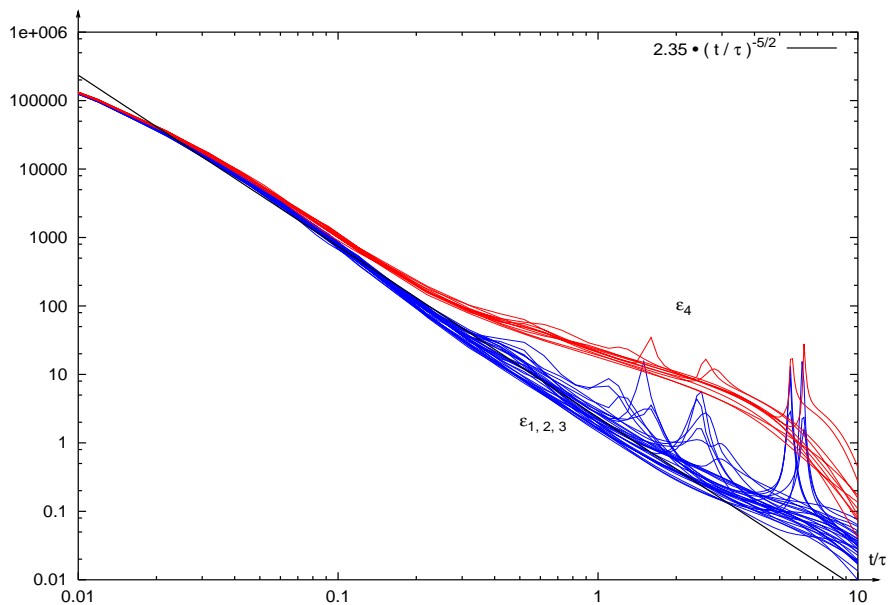}
\caption{\label{rates}
The rates of energy loss $\epsilon_a$ 
(\ref{epsi}) (for $a$=1,2,3 and $a=4$) through emission of 
pions and $\sigma$-mesons for a sample of 10 events evolving on an
$L=30$ cubic 
lattice after a sudden quench, with potential strength $\kappa^2$=20.
For comparison the power-law 2.35 $(t/\tau)^{-2.5}$ is indicated by the
straight line. } 
\end{figure}

Despite this significant growth of the width of abundance
ratio distributions for neutral pions, observable consequences are 
strongly suppressed by a rapid decrease of the absolute 
pion yield rates during the roll-down phase. As obtained 
according to eq.(\ref{epsi}) from the square of the instantaneous field
velocities the emission rates $\epsilon_a$ for $a$=1,2,3,4
are shown in Fig.(\ref{rates}) for the same sample as in
Fig.(\ref{width}).  
These rates decrease by three to four orders
of magnitude during the roll-down phase, whereby the explicit symmetry
breaking enhances the time-derivatives of the 4-component in such a way
that the time-integrated rates $\int \epsilon_4 dt$ are comparable
to the sum of the three pionic components $\sum_{i=1,2,3} \int
\epsilon_i dt$ (integrated over the roll-down phase). 
During this time interval the pionic emission rates (unaffected by
explicit symmetry breaking) decrease approximately like $ \sim
t^{-\beta}$ with $\beta\approx 2.3-2.5$. 

The observed total number $n$ of all pions emitted
in a single event is given by the time integral 
\be
\label{n}
n=\int_{t_0}^{\infty} {\dot n}(t) dt=
\int_{t_0}^{\infty} \sum_{i=1,2,3}\epsilon_i(t)/m_\pi dt.
\ee
where the initial time $t_0$ marks the onset of the emission. If we
approximate the instantaneous emission rate ${\dot n}(t)$ for pions 
emitted from a source of volume $V$ by a power law
${\dot n}(t)=\gamma V\: (t/t_0)^{-\beta}$, with constant emission rate
density $\gamma$, we have
\be
\label{n1}
n=\frac{\gamma V}{(\beta-1)}t_0.
\ee
If we choose for $t_0$ the onset of the
roll-down phase, we find for the total energy $E$ at that time
typically $E(t_0/\tau \sim 0.1) 
\approx 0.3{\mbox { GeV/fm}}^3~ (La)^3$. With lattice constant
$a=1/6$ fm and lattice size $L=60$ the total energy of about 300 GeV
residing at time $t_0$ in this (10 fm)$^3$ fireball 
would be dissipated through emission of about
150 GeV/$m_\pi \approx 1100$ low-energy pions and 150 GeV/$m_\sigma$
$\sigma$-mesons. With  $\beta=5/2$, we obtain for the initial
pion emission density at time $t_0/\tau=0.1$,  $\gamma \tau \approx 16$
fm$^{-3}$. Evidently, most of the observed pions are emitted at the very
beginning of the roll-down phase.

For a random (non-DCC) source the width of the
probability distribution for the fraction $f$ of neutral pions
is given by the binomial result
\be
\label{sig}
\sigma_{non-DCC}^2=\frac{2}{9n}
\ee
where the total number $n$ of pions emitted now is given by (\ref{n1}).
For a source with $\nu(t)$ DCC domains present at time $t$
the width of the
probability distribution for the observed fraction $f$ of all neutral pions
emitted in that event is (with (\ref{sigdcc}))
\be
\label{sigDCC}
\sigma_{DCC}^2=\frac{1}{n}\int_{t_0}^{\infty}
{\dot n}(t)\left(\frac{4}{45\nu(t)}\right) dt.
\ee
If the number $\nu(t)$ of DCC-domains present at time $t$ within the
volume $V$ 
is approximated by the power law $\nu(t)=\nu_0 \:(t/t_0)^{-3\alpha}$, 
where $\nu_0$ denotes the number of DCC-domains present at time $t_0$,
we have 
\be
\label{nueff}
\sigma_{DCC}^2= \frac{4}{45\: \nu_{eff}},~~~~~{\mbox{ with     }}
\nu_{eff}=
\left(\frac{\beta-3\alpha-1}{\beta-1}\right)\:\nu_0.
\ee
For positive values of $\alpha$ the effective number of DCC-domains
$\nu_{eff}$ is reduced as compared to the number $\nu_0$ of domains
present at the time $t_0$ of the onset of pion  emission. 
(Of course, we have assumed here that the Gaussian approximation
remains valid as long as there is noticeable emission strength 
${\dot n}(t)$. If, however, the growth rate $3\alpha$ of the average volume
of one DCC-domain is comparable to ($\beta-1$), the effective number 
$\nu_{eff}$ becomes very small, the Gauss approximation breaks down,
and we may expect the anomalies which characterize $P_\nu(f)$ for
very small values of $\nu$ as shown in Fig.(\ref{Pfig})).

The simulations indicate that $\beta\approx 2.3-2.5$ and $\alpha\approx
0.25-0.3$, so we find 
\be
\nu_{eff}/\nu_0=0.175-0.5
\ee
which implies a broadening of the distribution $P(f)$ by a factor
of $1.4-2.4$ as compared to the distribution at time $t_0$. It may be
noted that this result is quite sensitive to the actual values of
$\alpha$ and $\beta$: a growth rate of the angular correlation length
$R_\pi \sim t^{1/3}$ instead of $t^{1/4}$, and the
emission rates decreasing like $t^{-2}$ instead of $t^{-2.5}$, 
would lead to the 
limiting case $3\alpha = \beta-1$. Unfortunately, the result
(\ref{nueff}) provides only 
a relative statement between the observed width $\sigma_{DCC}$ and 
a hypothetical width at time $t_0$, and not a comparison with the
width (\ref{sig}) for a binomial distribution. We could enforce such a
connection by assuming that for $\alpha=0$ (zero growth rate for
DCC domains) the source should be undistinguishable from a randomly
emitting source. In that case
we would have 
\be
\sigma_{DCC}=\sigma_{non-DCC} \sqrt{\left(
\frac{\beta-1}{\beta-3\alpha-1}\right)}.
\ee
So, the growth of the DCC domains, in competition with the decrease
of the emission rates, still would lead to a significant broadening in the
effective distribution of $\pi_0$-abundance ratios, as compared to the
corresponding width (\ref{sig}) for a random non-DCC source. 

The growth law for the angular correlation length $R_\pi$
and the roll-down times $t_{RD}$ are not very sensitive to the 
(sufficiently large) total lattice size $L$, therefore the width
$\sigma_{DCC}$ of the abundance ratio distributions scales with
$L^{-3/2}$. (The same holds for the binomial distribution).
On the other hand, the amplitudes of pronounced peaks in the abundance 
ratios $f_0[\dot{\Phi}]$ due to restructuring textures, i.e. $B-{\bar
B}$ annihilation, are independent of $L$, while their number scales with
$L^3$. So in larger fireballs we would expect additional broadening of
the width 
of fluctuations around the value of 1/3 to be increasingly due to
annihilations of nascent baryon-antibaryon structures, without the
specific anomalies which characterize pion emission from large
disoriented domains.

\subsection{Slow cooling}

Up to now we have discussed the extreme case of a sudden quench where
the potential (\ref{pot}) at time $t=0$ 
drops instantly to its temperature $T=0$ shape, i.e. $f_0^2(T)=f_\pi^2$
for $t>0$. Practically, however, for a smooth quench it may take a
typical cooling time $\tau_c$ for 
$f_0^2(T)$ to approach its $T=0$ value $f_\pi^2$. As long as $\tau_c$
is small as compared to the typical relaxation times $\tau$ of
the system, the evolutions proceed much like those for a sudden quench.
Noticeable differences appear only if the cooling time $\tau_c$
becomes comparable to the typical roll-down time $t_{RD}$ following the
sudden quench. 

In addition to the smooth time-dependence of $f_0(T(t))$, the 
field configurations may be subjected to a stochastic force 
which describes the back-reaction of the eliminated fluctuations on the
slowly evolving degrees of freedom, acting as a heat bath for the 
cooling fireball. As we have
assumed in eq.(\ref{epsi}) that the main source of energy loss is through
pion and $\sigma$ emission, with the fireball immersed in the $T=0$ vacuum, 
the strength of the stochastic force is not strictly tied 
to the relaxation constant $\tau$, but will be comparatively small.
Let us therefore separately discuss the influence of the smoothly
time-varying coefficient $f^2$ in the $\Phi^4$-potential, without any
noise term.

The first phase (I) of the ordering evolution proceeds independently
from the strength of the $\Phi^4$ potential, so it is also insensitive
to the actual value of $f^2(T)$. The angular correlation lengths and
the baryon-plus-antibaryon densities established by the end of this
phase are independent of the cooling time $\tau_c$. 

Subsequently, during the roll-down phase (II) of a slow cooling process 
with $\tau_c \gg t_{RD}$, the configurations which emerge near the end of
phase (I) now have sufficient time to relax towards a
minimum at the instantaneous temperature.
As a result of the baryon radius $R$ scaling as $R \sim (e
f_0(T))^{-1}$, with $f_0(T)/f_\pi < 1$ throughout 
most of the interior volume, the nascent textures are
slightly larger, overlap and interact more strongly, and
therefore annihilate or fuse more easily into multi-$B$ configurations.
All these textures are well developed by the end of phase (II) although
the actual condensate $\langle \Phi \rangle$ is close to the 
momentaneous value of $f_0(T(t))$, i.e. much smaller than $f_\pi$. 

During the following increase of $\langle \Phi \rangle$ towards $f_\pi$
which proceeds on the time scale $\tau_c$, the established
configurations simply follow the changing effective potential in a
more or less equilibrated manner. The addition of a small temperature-dependent
noise term causes only a slight retardation in the ordering evolutions.

Altogether, the {\it angular} ordering proceeds on the time scale of the
relaxation constant $\tau$, quite independently of the cooling
time scale $\tau_c$. From the point of DCC effects, the case of the
sudden quench shows the essential features. The late-time reshaping
of the established textures according to a changing $f_0(T(t))$ is 
not relevant for the observability of DCC pion anomalies.

\section{Conclusion}

As a dynamical model for the relaxation of a hot chirally symmetric
'fireball' towards the cold vacuum characterized by 
spontaneously broken chiral symmetry with nonvanishing chiral condensate,
we have investigated the classical time evolution of chiral $O$(4)-field
configurations. Starting from stochastic initial conditions, which are
chirally symmetric in the interior of a finite spatial volume, the
evolutions of the classical fields follow purely dissipative 
dynamics in their relaxation towards the physical vacuum which
surrounds the considered volume as boundary condition.

The model is limited by a severe restriction: The chiral $O$(4)-field
comprises all degrees of freedom of the relaxing fireball. This renders
it questionable that the model can be applied and trusted at conditions
which may prevail at temperatures close to or above the critical $T_c$
of the chiral phase transition. Consequently, we do not pay much
attention to details of the initial conditions. The
dissipative dynamics wipes them out, anyway, after a small fraction of the 
typical relaxation time unit. But we hope, that once the color degrees
of freedom are \mbox{(re-)} confined into the hadronic chiral field,
and the energy-density has dropped to less than 1GeV/fm$^3$, 
then the model may provide a reliable picture for the later stages of the
evolution when the actual roll-down of the order parameter takes place. 

While the dissipative term during the very early phase of the
evolution may originate from the rapid expansion of the hot
fireball, we assume that during the subsequent roll-down phase 
energy is lost and carried
away by emission of chiral field quanta (pions and $\sigma$-mesons).
Then the square of the instantaneous local field velocities provide a
measure for the strength of the emitted radiation. 
Two features are of specific interest: 
The growth and typical size of domains with disoriented
chiral field, and the possibility to observe anomalies in the abundance
ratios of emitted pions caused by the transient formation of such
domains. 

On the other hand, the model is sufficiently sophisticated to allow for
creation, stabilization and mutual interactions of baryons and
antibaryons in the hot hadronic gas, and their presence as extended
topological textures is of specific importance for the ordering process.  
We consider it as one of the essentials of the present investigation to use
an effective action which incorporates the possibility to create
and stabilize these structures, because it is well known that textures 
are of crucial importance for the phenomenology of phase transitions.
Their physical size and properties set the scale for the stabilizing
and for the explicitly symmetry-breaking terms in the effective action.
Numerical implementation on a 3+1 dimensional lattice then requires a
lattice constant $a$ which allows to describe these textures with
appropriate accuracy. It turns out that \mbox{$a$=1/6 fm} is sufficient for
that purpose. Typical fireball diameters in heavy-ion collisions in the
range from 5 to 15 fm then require lattice sizes of about \mbox{$30<L<90$},
which are numerically reasonable to handle.

The essentials of our results are the following:

During the very early stages of the evolutions the 
angular gradient terms which drive the local aligning of the field
vectors dominate over the $\Phi^4$ potential such that the actual
roll-down of the modulus $\Phi$ towards its vacuum value sets in only
after angular correlation lengths of about 0.5-0.7 fm have been established
(see Fig.(\ref{field}c), and see the angular correlation radii $R_\pi$ in
Fig.(\ref{rd30}) near $t/\tau$=0.1-0.2). These configurations set the stage
for the subsequent roll-down.

During the roll-down phase the angular ordering 
is dominated by the higher-order gradient terms in the effective action
which try to establish low-energy configurations for the emerging
extended textures. This leads to an appreciable reduction in the
growth exponent $\alpha$ of the angular correlation length $R_\pi$,
which measures the size of aligned disoriented domains in the isospin
subspace, i.e. orthogonal to the 4-direction. 
Due to this slow growth, the correlation lengths $R_\pi$ reach values
of only up to about 10 lattice units by the end of the roll-down phase
(depending on the number of remaining baryons) which corresponds to
disoriented domain radii of about 1.5 fm, independently of the total
fireball size. A nonvanishing number of baryons (plus antibaryons)
which by that time still may be floating within the developing vacuum
configuration, causes the radii $R_\pi$ to saturate at even smaller values.
This result, that near the end of the roll-down the disoriented domains
are essentially pion-sized has previously been suggested by 
qualitatively quite different arguments~\cite{Gavin}. 
It is interesting that the purely
dissipative dynamics used here, apparently leads to a similar
conclusion. The result implies that even for the smallest fireball
considered (with radius $aL/2$=2.5 fm) we expect more than at least 5
distinguishable domains to be present at the end of the roll-down phase.

The realistic strength of explicit symmetry breaking in 4-direction, 
determined by the physical value of the ($T$=0) pion mass and by the 
influence of the ($T$=0) vacuum surrounding the hot finite volume,
is sufficiently strong to effectuate saturation of the order parameter
$\langle \Phi \rangle$ by the average 4-component $\langle \Phi_4
\rangle$ even before the order parameter has reached its vacuum value.
In other words, alignment in vacuum direction is completed before the
roll-down is complete. This not only holds for the lower limit
of the $\sigma$-mass $m_\sigma/f_\pi \approx 6$ ( weak
$\Phi^4$-potential), but still is basically true also for 
$m_\sigma/f_\pi \approx 10$ (strong $\Phi^4$-potential).

The number $N$ of baryons plus antibaryons still present
near the end of the roll-down is quite small, 
corresponding to densities of about 0.02/fm$^3$, with large
event-by-event fluctuations. Their number further decreases after
completion of the roll-down due to later annihilations. 
Each one carries a typical energy of $E_B=6\pi^2 f_\pi/e \approx 1.3$
GeV. So, during the roll-down phase the 
energy density drops from about 0.5 GeV/fm$^3$ to almost zero. In our picture
all of that is carried away by low-energy pions and $\sigma$'s with
the emission rates $\epsilon_a$ driven by the squares of the field
velocities (\ref{epsi}).  These emission rates drop rapidly during
roll-down like $\sim t^{-\beta}$ with $\beta \approx 2.3-2.5$ 
such that near $t/\tau\approx 1$, already $90\%$ of the
roll-down pions have been emitted. 
At that time, the radii of disoriented aligned domains have grown to
at most 1 fm (6 lattice units), and the three pionic rates $\epsilon_i$
are still degenerate with good accuracy. In accordance with the growth
of the angular correlation lengths like $\sim t^\alpha$ with $\alpha
\approx 0.25-0.3$ the 
variance in the fluctuations of the abundancy ratio for neutral pions
(obtained from the field velocities) grows like $\sim t^{3\alpha}$.  
Later on during the evolution occasional strong deviations from
degeneracy are due to changes in the structure of the emerging
textures, rather than to reorientations of aligned domains. 

For the sizes of fireballs considered here (with radii larger than about 5
fm), during that early phase of the roll-down when most of the pions are
emitted, the number of DCC domains present is sufficiently large to prevent
significant deviations from the shape of a Gaussian
distribution in the charge fluctuations. However, their transient
formation during the chiral phase transition would lead to a broadening
of the width of this distribution by a factor of
$\sqrt{(\beta-1)/(\beta-3\alpha-1)}$ as compared to a randomly emitting
uncorrelated source. For the values of $\alpha$ and $\beta$ extracted
from our simulations this factor amounts to at least $\sim \sqrt{2}$. Slightly
smaller values for $\beta$ and larger values for $\alpha$ which might
still be compatible with the simulations could increase this factor
appreciably. The slow growth of the angular correlation lengths which
is reflected in the smallness of $\alpha$ is due to the presence
of stabilizing extended textures within the ordering chiral field
which constitutes an essential feature of the present investigation.

So, if the present model with its underlying dissipative dynamics
indeed provides a reasonably realistic description for the roll-down
of a hot symmetric configuration into the spontaneously symmetry-broken
$T=0$ vacuum, then we would finally conclude that the transient
formation of DCC domains  would
hardly cause significant anomalous deviations from the Gaussian shape
for the distributions of neutral-to-charged ratio fluctuations. This
seems to be in accordance with the present experimental
situation~\cite{Exp}. The model simulations, however, could lead to the
expectation that the relaxation to the true vacuum after a chiral
phase transition should manifest itself in a broadening of the width 
of the distribution as
compared to standard statistical expectations, by a factor of
about $1.4-2.4$.

\acknowledgements
The authors appreciate numerous helpful discussions with H. Walliser.

\end{document}